%% file: 21_10_29_1bitADCs.tex
\documentclass[journal,romanappendices]{IEEEtran}

\IEEEoverridecommandlockouts

\usepackage{algorithm,algorithmic,amsmath,amssymb,amsthm,bbm,caption,cite,color,comment,graphicx,microtype,setspace,subcaption,tabularx,url}
\usepackage[USenglish]{babel}
\usepackage[T1]{fontenc}
\usepackage[utf8]{inputenc}
\usepackage{hyperref}

\usepackage{enumitem}
\setitemize{itemsep=0mm,topsep=0mm,parsep=0pt,partopsep=0pt}

\usepackage{tikz}
\usetikzlibrary{arrows,backgrounds,calc,intersections,decorations.pathreplacing,positioning,shapes}

\makeatletter
\newcommand\subparagraph{%
  \@startsection{subparagraph}{5}
  {\parindent}
  {3.25ex \@plus 1ex \@minus .2ex}
  {-1em}
  {\normalfont\normalsize\bfseries}}
\makeatother
\usepackage{titlesec}
\let\subparagraph\relax

\usepackage{titlesec}
\let\subparagraph\relax
\titlespacing{\section}{0pt}{5pt plus 2pt minus 1pt}{3pt plus 1pt minus 0pt}
\titlespacing{\subsection}{0pt}{4pt plus 2pt minus 1pt}{2pt plus 1pt minus 0pt}
\usepackage[figurename=Fig.,font=small]{caption}

{\newtheorem{theorem}{Theorem}}
{}
{}
{\newtheorem{proposition}{Proposition}}
{}
{}
{\newtheorem{corollary}{Corollary}}

\input{./Definitions.tex}
\newcommand{\blm}{\textrm{BLM}}
\newcommand{\El}{\mathsf{E}_{\ell}}
\newcommand{\erf}{\mathrm{erf}}
\newcommand{\mse}{\textrm{MSE}}
\newcommand{\rmp}{\textrm{p}}

\newcommand{\sls}{\textrm{SLS}}
\newcommand{\Vl}{\mathsf{V}_{\ell}}

\title{Channel Estimation and Data Detection \\ Analysis of Massive MIMO with 1-Bit ADCs}
\author{Italo Atzeni,~\IEEEmembership{Member,~IEEE}, and Antti Tölli,~\IEEEmembership{Senior~Member,~IEEE}
\thanks{The authors are with the Centre for Wireless Communications, University of Oulu, Finland (emails: \{italo.atzeni, antti.tolli\}@oulu.fi). The work of I.~Atzeni was supported by the Marie Sk\l{}odowska-Curie Actions (MSCA-IF 897938 DELIGHT). The work of A.~Tölli was supported by the Academy of Finland (318927 6Genesis Flagship and 319059 CCCWEE). Part of this work has been presented at IEEE SPAWC 2021, Lucca, Italy, Sept. 2021 \cite{Atz21}.}}

\begin{document}

\maketitle

\begin{abstract}
We present an analytical framework for the channel estimation and the data detection in massive multiple-input multiple-output uplink systems with 1-bit analog-to-digital converters (ADCs) and i.i.d. Rayleigh fading. First, we provide closed-form expressions of the mean squared error (MSE) of the channel estimation considering the state-of-the-art linear minimum MSE estimator and the class of scaled least-squares estimators. For the data detection, we provide closed-form expressions of the expected value and the variance of the estimated symbols when maximum ratio combining is adopted, which can be exploited to efficiently implement minimum distance detection and, potentially, to design the set of transmit symbols. Our analytical findings explicitly depend on key system parameters such as the signal-to-noise ratio (SNR), the number of user equipments, and the pilot length, thus enabling a precise characterization of the performance of the channel estimation and the data detection with 1-bit ADCs. The proposed analysis highlights a fundamental SNR trade-off, according to which operating at the right noise level significantly enhances the system performance.

\end{abstract}

\section{Introduction} \label{sec:INTRO}

The migration of operating frequencies from first- to fourth-generation wireless systems, i.e., from 800~MHz to the sub-3~GHz range, did not bring major changes in terms of signal propagation. The current fifth generation (5G) features a more pronounced transition in this respect by operating at sub-6~GHz frequencies and, eventually, up to 30~GHz with the objective of boosting the data rates. Following this trend, beyond-5G systems will exploit the large amount of bandwidth available in the mmWave band (i.e., 30~GHz--300~GHz) and raise the operating frequencies up to 1~THz \cite{Raj20}. In this context, maintaining the same signal-to-noise ratio (SNR) over a given distance will require larger antenna arrays and increasingly sharp beamforming to spatially focus the signal power. Although the short wavelength at mmWave and sub-THz frequencies allows to pack many antennas into a very small area, realizing fully digital, high-resolution massive multiple-input multiple-output (MIMO) arrays remains prohibitive in practice \cite{Rus13,Xia17}.

\begin{figure}[t!]
\centering
\includegraphics[scale=1]{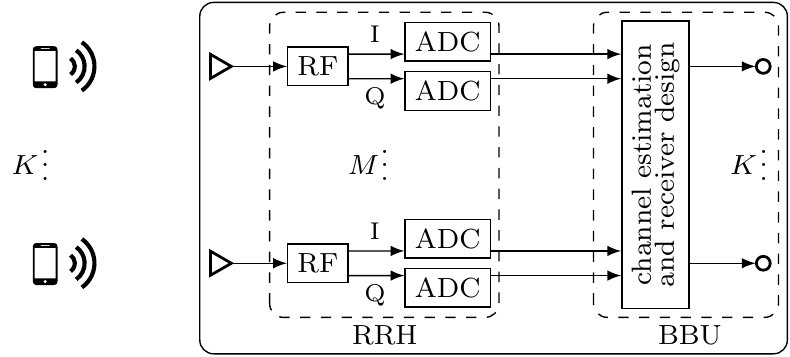} \vspace{2mm}
\caption{Fully digital massive MIMO uplink system.} \label{fig:SM} \vspace{-4mm}
\end{figure}

As in the system model illustrated in Fig.~\ref{fig:SM}, each base station (BS) antenna is generally equipped with a dedicated radio-frequency (RF) chain that includes complex, power-hungry analog-to-digital/digital-to-analog converters (ADCs/DACs) \cite{Xia17}. In this setting, while the transmit power can be made inversely proportional to the number of antennas, the power consumed by each ADC/DAC scales linearly with the sampling rate and exponentially with the number of quantization bits \cite{Mo15,Stu16,Cho16,Li17,Sax17}. Another limiting factor is the amount of raw data exchanged between the remote radio head (RRH) and the base-band unit (BBU), which scales linearly with both the sampling rate and the number of quantization bits \cite{Jac17,Jac17a,Jac19}. For these reasons, adopting low-resolution ADCs/DACs with 1 to 4 quantization bits as opposed to the typical 10 or more \cite{Sin09} enables the implementation of massive MIMO arrays comprising hundreds (or even thousands) of antennas, which are necessary to operate in the mmWave and sub-THz bands \cite{Jac17}. In this regard, 1-bit ADCs/DACs are particularly appealing due to their minimal power consumption and complexity since they only evaluate the sign of the input signal \cite{Mo15}. Such a coarse quantization is especially motivated at very high frequencies, where high-order modulations are not essential.

There is a large body of literature on massive MIMO with low-resolution and 1-bit ADCs/DACs, ranging from performance analysis to channel estimation and precoding design. The capacity of the low-resolution and the 1-bit quantized MIMO channel is characterized in \cite{Mez12} and \cite{Mo15}, respectively, whereas \cite{Lia16} shows that replacing even a small number of high-resolution ADCs with 1-bit ADCs entails a modest performance loss. The performance-quantization trade-off in orthogonal frequency-division multiplexing (OFDM) uplink systems is studied in \cite{Stu16}, which shows that using 4 to 6 quantization bits involves almost no performance loss compared with infinite-resolution ADCs. The spectral efficiency of single-carrier and OFDM uplink systems with 1-bit ADCs is analyzed in \cite{Mol17}. The problem of multi-user detection is considered, e.g., in \cite{Wan15} and \cite{Jeo18} for low-resolution and 1-bit ADCs, respectively, whereas \cite{Wen16} focuses on the joint channel estimation and data detection. An efficient iterative method for near maximum likelihood detection with 1-bit ADCs is proposed in \cite{Cho16}. The work in \cite{Li17} analyzes the channel estimation and the uplink achievable rate with 1-bit ADCs. In addition, it proposes a linear minimum mean squared error (MMSE) channel estimator based on the Bussgang decomposition, which allows to reformulate the nonlinear quantization function as a linear function with identical first- and second-order statistics \cite{Bus52}: we refer to this estimator as Bussgang linear MMSE (BLM) estimator. A similar analysis is presented in \cite{Li17a} for the downlink direction. The work in \cite{Jac17} extends some of the results derived in \cite{Mol17,Li17} for 1-bit ADCs to the multi-bit case. Specifically, it presents a throughput analysis of uplink systems and proposes a linear channel estimator based on the Bussgang decomposition with low-resolution ADCs. The channel estimation with 1-bit ADCs when the quantization threshold is not known is studied in \cite{Ste18}. The channel estimation exploiting the angular and delay structure is considered in \cite{Mo18} and \cite{Rao19} for low-resolution and 1-bit ADCs, respectively, whereas \cite{Kim19} exploits the temporal correlation for 1-bit ADCs. A recent line of works employs machine learning techniques in scenarios where obtaining accurate channel state information with low-resolution ADCs is impractical (see, e.g., \cite{Jeo20,Ngu20}). The benefits of oversampling for 1-bit quantized uplink systems are investigated in \cite{Ucu18,Sha20}. The performance of linear precoding schemes for downlink systems with 1-bit DACs is analyzed in \cite{Sax17}. A similar analysis with multi-bit DACs is presented in \cite{Jac17a} considering both linear and nonlinear precoding, and in \cite{Jac19} considering linear precoding with oversampling in OFDM downlink systems. Lastly, \cite{Sha19} proposes a general optimization framework for downlink precoding with 1-bit DACs and constant envelope assuming quadrature amplitude modulation (QAM) transmit symbols.

\subsection{Contribution}

This paper broadens prior analytical studies on the channel estimation and the data detection in massive MIMO uplink systems with 1-bit ADCs. On the one hand, existing works do not provide a precise characterization of the performance of the channel estimation with 1-bit ADCs with respect to key system parameters such as the SNR, the number of user equipments (UEs), and the pilot length. We fill this gap by analyzing the mean squared error (MSE) of the channel estimation along with its asymptotic behavior at high SNR assuming independent and identically distributed (i.i.d.) Rayleigh fading channels among the UEs. In this respect, we consider the BLM estimator in \cite{Li17} as well as the class of scaled least-squares (LS) estimators, such as the one proposed in \cite{Mol17} (which can be obtained from the former by ignoring the temporal correlation of the quantization distortion). On the other hand, in the context of data detection with 1-bit ADCs, the statistical properties of the estimated symbols have not been characterized by existing works. In this regard, an interesting SNR trade-off was observed in \cite{Jac17}, whereby the estimated symbols resulting from transmit symbols with the same phase overlap at high SNR; however, this aspect has not been formally described in the literature. We fill this gap by analyzing the expected value and the variance of the estimated symbols along with their asymptotic behavior at high SNR.\footnote{This second part of the paper complements the results of \cite{Atz21} with a detailed analysis of the normalized variance of the estimated symbols (also by means of tractable upper bounds), while the first part is entirely new.} Our results on both the channel estimation and the data detection ultimately impact the symbol error rate (SER) and thus provide important practical insights into the design and the implementation of 1-bit quantized systems.

The contributions of this paper are summarized as follows:
\begin{itemize}
\item[$\bullet$] For the channel estimation with 1-bit ADCs, we derive closed-form expressions for the BLM estimator in \cite{Li17} and for the class of scaled LS estimators, such as the one proposed in \cite{Mol17}. This enables a precise characterization of the performance of the channel estimation with respect to the SNR, the number of UEs, and the pilot length. Furthermore, we show that, in the case of i.i.d. Rayleigh fading channels among the UEs, the BLM estimator can be simplified as a scaled LS estimator with UE-specific scaling factors and that using a common optimized scaling factor for all the UEs entails a negligible performance loss.
\item[$\bullet$] For the data detection with 1-bit ADCs, we characterize the statistical properties of the estimated symbols by deriving closed-form expressions of the expected value and the variance when maximum ratio combining (MRC) is adopted at the BS. These results can be exploited to efficiently implement minimum distance detection (MDD) and, potentially, to design the set of transmit symbols to further improve the data detection performance.
\item[$\bullet$] Building on the proposed analysis, we provide a thorough discussion on the effect of 1-bit quantization on both the channel estimation and the data detection. For each of the two aspects, we describe a fundamental SNR trade-off, according to which operating at the right noise level significantly enhances the system performance. In this respect, the optimal transmit SNR for the channel estimation is shown to decrease as the pilot length increases.
\end{itemize}

\vspace{1mm}

\textit{Outline.} The rest of the paper is structured as follows. Section~\ref{sec:SM} introduces the system model with 1-bit ADCs. Sections~\ref{sec:PA_EST} and~\ref{sec:PA_DEC} present our performance analysis results on the channel estimation and the data detection, respectively, each including dedicated numerical results and discussions. Finally, Section~\ref{sec:CONCL} summarizes our contributions and draws some concluding remarks.

\vspace{1mm}

\textit{Notation.} $\A = (A_{m,n})$ specifies that $A_{m,n}$ is the $(m,n)$th entry of matrix $\A$; likewise, $\a = (a_{n})$ specifies that $a_{n}$ is the $n$th entry of vector $\a$. $(\cdot)^{\tran}$, $(\cdot)^{\herm}$, and $(\cdot)^{*}$ represent the transpose, Hermitian transpose, and conjugate operators, respectively. $\Re[\cdot]$ and $\Im[\cdot]$ denote the real part and imaginary part operators, respectively, whereas $j$ is the imaginary unit. $\Exp [\cdot]$ and $\Var [\cdot]$ are the expectation and variance operators, respectively. $\I_{N}$, $\0_{N}$, and $\1_{N}$ denote the $N$-dimensional identity matrix, all-zero vector, and all-one vector, respectively. $\Diag(\cdot)$ produces a diagonal matrix with the entries of the vector argument on the diagonal, $\sgn(\cdot)$ is the sign function, and $\mathrm{vec}[\cdot]$ is the vectorization operator. The Kronecker product is denoted by $\otimes$ and $\{\cdot\}$ is used to represent sets. Lastly, $\setC \setN(0, 1)$ is the complex normal distribution with zero mean and unit variance, whereas $\setN(\0_{N}, \Sigmab)$ is the real $N$-variate normal distributions with zero mean and covariance matrix $\Sigmab$.

\vspace{1mm}

\textit{Reproducible research.} All the numerical results can be reproduced using the MATLAB code and data files available at: \href{https://github.com/italo-atzeni/Ch_Est_Data_Det_1-Bit}{https://github.com/italo-atzeni/Ch\textunderscore Est\textunderscore Data\textunderscore Det\textunderscore 1-Bit}.

\section{System Model} \label{sec:SM}

Consider the scenario depicted in Fig.~\ref{fig:SM}, where a BS with $M$ antennas serves $K$ single-antenna UEs in the uplink. Let $\H \triangleq (H_{m,k}) \in \Compl^{M \times K}$ denote the uplink channel matrix: assuming i.i.d. Rayleigh fading (as, e.g., in \cite{Jac17,Li17,Jac17a,Jac19}), the entries of $\H$ are distributed independently as $\setC \setN (0,1)$. Each UE transmits with power $\rho$ and the additive white Gaussian noise (AWGN) at the BS has unit variance: thus, $\rho$ can be interpreted as the transmit SNR. Note that the same transmit SNR is assumed for the two phases of channel estimation and uplink data transmission. Each BS antenna is connected to two 1-bit ADCs, one for the in-phase component and one for the quadrature component of the receive signal. Therefore, according to \cite{Jac17}, we introduce the 1-bit quantization function $Q(\cdot) : \Compl^{L \times N} \to \setQ$, with
\begin{align}
Q(\A) \triangleq \sqrt{\frac{\rho K + 1}{2}} \Big( \sgn \big( \Re[\A] \big) + j \, \sgn \big( \Im[\A] \big) \Big)
\end{align}
and where $\setQ \triangleq \sqrt{\frac{\rho K + 1}{2}} \{\pm 1 \pm j\}^{L \times N}$ is the set containing the scaled symbols of the quadrature phase-shift keying (QPSK) constellation.

\subsection{Channel Estimation} \label{sec:SM_est}

In the channel estimation phase, the UEs simultaneously transmit their uplink pilots of length $\tau$. Let $\P \triangleq (P_{u,k}) \in \Compl^{\tau \times K}$ denote the pilot matrix whose columns correspond to the pilots used by the UEs, with $|P_{u,k}|^{2} = 1$, $\forall u,k$. We assume $\tau \geq K$ and orthogonal pilots among the UEs, so that $\P^{\herm} \P = \tau \I_{K}$. Hence, the receive signal at the BS prior to quantization is given by
\begin{align}
\Y_{\rmp} \triangleq \sqrt{\rho} \H \P^{\herm} + \Z_{\rmp} \in \Compl^{M \times \tau}
\end{align}
where $\Z_{\rmp} \triangleq (Z_{m,u}) \in \Compl^{M \times \tau}$ is the AWGN term with entries distributed as $\setC \setN (0,1)$. Then, at the output of the ADCs, we have
\begin{align} \label{eq:R_p}
\R_{\rmp} & \triangleq Q(\Y_{\rmp}) \in \Compl^{M \times \tau}
\end{align}
with $\R_{\rmp} = (R_{m,u})$, which is used by the BS to estimate $\H$. Some comments are in order. First, correlating the quantized receive signal $\R_{\rmp}$ in \eqref{eq:R_p} with $\P$, as done in \eqref{eq:H_hat_sls} below, results in residual pilot contamination even when the pilots are orthogonal (see, e.g., \cite{Mol17}). Second, the pilots should be preferably chosen such that their entries span an interval $\big[ \eta, \eta + \frac{\pi}{2} \big]$, with $\eta \in [0, 2 \pi]$, so as to accurately estimate the phases (especially at high SNR). This is explained in Appendix~\ref{app:ch_est}, which provides a detailed discussion on the channel estimation with 1-bit ADCs.

The LS estimator for 1-bit ADCs, which correlates the quantized receive signal $\R_{\rmp}$ in \eqref{eq:R_p} with $\P$, was first presented in \cite{Ris14}. Then, a linear MMSE estimator based on the Bussgang decomposition (see \cite{Bus52}), which we refer to as BLM estimator, was proposed in \cite{Li17}. According to this, $\underline{\h} \triangleq \mathrm{vec}[\H] \in \Compl^{M K \times 1}$ is estimated as\footnote{Note that, in the case of correlated channels, the BLM estimator in \eqref{eq:H_hat_blm} is given by \cite[Eq.~(60)]{Li17} and the channel covariance matrix is also embedded into $\Sigmab_{\rmp}$.}
\begin{align} \label{eq:H_hat_blm}
\underline{\hat{\h}}_{\blm} \triangleq \sqrt{\frac{2}{\pi} \rho} \tilde{\P}^{\tran} \Sigmab_{\rmp}^{-1} \underline{\r}_{\rmp} \in \Compl^{M K \times 1}
\end{align}
with $\tilde{\P} \triangleq \P \otimes \I_{M} \in \Compl^{M \tau \times M K}$, $\underline{\r}_{\rmp} \triangleq \mathrm{vec}[\R_{\rmp}] \in \Compl^{M \tau \times 1}$, and where $\Sigmab_{\rmp} \triangleq \Exp [\underline{\r}_{\rmp} \underline{\r}_{\rmp}^{\herm}] \in \Compl^{M \tau \times M \tau}$ denotes the covariance matrix of $\underline{\r}_{\rmp}$. A linear estimator with a simpler structure can be obtained from \eqref{eq:H_hat_blm} by ignoring the temporal correlation of the quantization distortion, which implies that the off-diagonal entries of $\Sigmab_{\rmp}$ are zero. Such a scaled LS estimator was proposed in \cite{Mol17} (and later extended to the case of multi-bit ADCs in \cite{Jac17}), whereby $\H$ is estimated as
\begin{align} \label{eq:H_hat_sls}
\hat{\H}_{\sls} \triangleq \sqrt{\Psi} \R_{\rmp} \P \in \Compl^{M \times K}
\end{align}
where the scaling factor $\Psi$ (common for all the UEs) is defined as 
\begin{align} \label{eq:Psi}
\Psi \triangleq \frac{2}{\pi} \rho \bigg( \frac{2}{\pi} \rho (\tau - K) + \rho K + 1 \bigg)^{-2}.
\end{align}
We point out that $\Psi$ in \eqref{eq:Psi} implicitly depends on the channel distribution. Note that, when $\tau = K$, \eqref{eq:H_hat_sls} coincides with \eqref{eq:H_hat_blm} since $\Sigmab_{\rmp} = (\rho K + 1) \I_{M K}$; otherwise, \eqref{eq:H_hat_sls} accurately approximates \eqref{eq:H_hat_blm} at low SNR or when $K$ is large \cite{Jac17}. In Section~\ref{sec:PA_EST}, we analyze the performance of the BLM estimator in \eqref{eq:H_hat_blm} and of the class of scaled LS estimators, such as the one in \eqref{eq:H_hat_sls}. Moreover, we highlight the relationship between these two in the case of i.i.d. Rayleigh fading channels among the UEs.

\subsection{Uplink Data Transmission} \label{sec:SM_dec}

Let $x_{k} \in \Compl$ be the transmit symbol of UE~$k$, with $\Exp \big[ |x_{k}|^{2} \big] = 1$ and $\x \triangleq (x_{k}) \in \Compl^{K \times 1}$. The receive signal at the BS prior to quantization is given by
\begin{align}
\y \triangleq \sqrt{\rho} \H \x + \z \in \Compl^{M \times 1}
\end{align}
where
$\z \triangleq (z_{m}) \in \Compl^{M \times 1}$ is the AWGN term with entries distributed as $\setC \setN (0,1)$. Then, at the output of the ADCs, we have
\begin{align} \label{eq:r}
\r \triangleq Q(\y) \in \Compl^{M \times 1}
\end{align}
and the BS obtains a soft estimate of $\x$ as
\begin{align} \label{eq:x_hat}
\hat{\x} \triangleq \V^{\herm} \r \in \Compl^{K \times 1}
\end{align}
where $\V \in \Compl^{M \times K}$ is the combining matrix adopted at the BS. Finally, the data detection process associates each estimated symbol to a transmit symbol, e.g., via MDD. In Section~\ref{sec:PA_DEC}, we focus on characterizing the statistical properties of the estimated symbols when MRC is adopted at the BS.

\section{Channel Estimation with 1-bit ADCs} \label{sec:PA_EST}

In this section, we are interested in characterizing the performance of the channel estimation with respect to the different parameters when 1-bit ADCs are adopted at each BS antenna (see Section~\ref{sec:SM_est}). In doing so, we consider the BLM estimator in \eqref{eq:H_hat_blm} and the class of scaled LS estimators, such as the one in \eqref{eq:H_hat_sls}.

\subsection{MSE of the Channel Estimation} \label{sec:PA_EST_main}

The (normalized) MSE of the channel estimation when the BLM estimator is used is given by
\begin{align} \label{eq:MSE_blm_0}
\mse_{\blm} \triangleq \frac{1}{M K} \Exp \big[ \| \underline{\hat{\h}}_{\blm} - \underline{\h} \|^{2} \big]
\end{align}
with $\underline{\hat{\h}}_{\blm}$ defined in \eqref{eq:H_hat_blm}. In \cite[Eq.~(17)]{Li17}, the closed-form expression of \eqref{eq:MSE_blm_0} was derived for the case of $\tau = K$, which gives
\begin{align} \label{eq:MSE_blm_tau=K}
\mse_{\blm} & = 1 - \frac{2}{\pi} \frac{\rho K}{\rho K + 1}.
\end{align}
Note that the above expression is lower bounded by $1 - \frac{2}{\pi} \simeq 0.363$, which is achieved in the limit of $\rho K \to \infty$. Hence, in realistic scenarios (especially for small values of $K$), using a pilot length that is equal to the number of UEs results in quite inaccurate channel estimates. In general, $\tau$ should be sufficiently large to compensate for the low granularity of the ADCs, as detailed in Appendix~\ref{app:ch_est}. For this reason, we assume that the pilot matrix $\P$ is chosen such that $\P \P^{\herm}$ is circulant\footnote{For example, this condition is satisfied when $\P$ is a partial discrete Fourier transform (DFT) matrix, i.e., $\P$ is composed of any $K$ columns of the $\tau$-dimensional DFT matrix.} and derive a closed-form expression of \eqref{eq:MSE_blm_0} that is valid for any value of $\tau$ and $K$.

\begin{theorem} \label{thm:MSE_blm} \rm{
Suppose that the BLM estimator in \eqref{eq:H_hat_blm} is used and that $\P$ is chosen such that $\P \P^{\herm}$ is circulant. Then, the MSE of the channel estimation in \eqref{eq:MSE_blm_0} is given by
\begin{align} \label{eq:MSE_blm}
\mse_{\blm} = 1 - \frac{2}{\pi} \frac{\rho \tau^{2}}{\rho K + 1} \frac{1}{K} \sum_{k=1}^{K} \frac{1}{\tau + \delta_{k}}
\end{align}
where we have defined
\begin{align}
\nonumber \delta_k & \triangleq \sum_{u \neq v} \bigg( \Re[P_{u,k}^{*} P_{v,k}] \Omega \bigg( \frac{\rho \sum_{i=1}^{K} \Re[P_{u,i} P_{v,i}^{*}]}{\rho K + 1} \bigg) \\
\label{eq:delta_k} & \phantom{=} \ - \Im[P_{u,k}^{*} P_{v,k}] \Omega \bigg( \frac{\rho \sum_{i=1}^{K} \Im[P_{u,i} P_{v,i}^{*}]}{\rho K + 1} \bigg) \bigg)
\end{align}
with
\begin{align}
\Omega(w) \triangleq \frac{2}{\pi} \arcsin(w).
\end{align}}
\end{theorem}

\begin{IEEEproof}
See Appendix~\ref{app:MSE_blm}.
\end{IEEEproof} \vspace{1mm}

\noindent The result of Theorem~\ref{thm:MSE_blm} enables a precise characterization of the performance of the BLM estimator with respect to the transmit SNR $\rho$, the number of UEs $K$, and the pilot length $\tau$. The parameter $\delta_{k}$ in \eqref{eq:delta_k} is roughly proportional to $\tau (\tau - 1)$ and, for a fixed $\tau$, decreases with $K$. When $\tau = K$, \eqref{eq:MSE_blm} recovers the expression in \eqref{eq:MSE_blm_tau=K}: in fact, $\tau = K$ implies $\delta_{k} = 0$, $\forall k$, since $\sum_{i=1}^{K} P_{u,i} P_{v,i}^{*} = 0$, $\forall u \neq v$. Moreover, the choice of the pilot matrix $\P$ affects the MSE of the channel estimation through the parameters $\{\delta_{1}, \ldots, \delta_{K}\}$. Lastly, we point out that, if $\P \P^{\herm}$ is not circulant, $\mse_{\blm}$ can be computed via the more involved and less insightful expression in \eqref{eq:MSEp_blm1} (see Appendix~\ref{app:MSE_blm}).

We now show that, in the case of i.i.d. Rayleigh fading channels among the UEs and when $\P \P^{\herm}$ is circulant, the BLM estimator can be simplified as a scaled LS estimator with UE-specific scaling factors.

\begin{corollary} \label{cor:MSE_blm_sls} \rm{
Suppose that $\P$ is chosen such that $\P \P^{\herm}$ is circulant. Then, the BLM estimator in \eqref{eq:H_hat_blm} can be simplified as
\begin{align} \label{eq:H_hat_blm_sls}
\hat{\H}_{\blm} \triangleq \R_{\rmp} \P \Psib^{\frac{1}{2}} \in \Compl^{M \times K}
\end{align}
where we have defined $\Psib \triangleq \Diag \big( [\Psi_{1}, \ldots, \Psi_{K}] \big) \in \Real^{K \times K}$, with
\begin{align} \label{eq:Psi_k}
\Psi_{k} \triangleq \frac{2}{\pi} \frac{\rho \tau^{2}}{(\rho K + 1)^{2} (\tau + \delta_{k})^{2}}.
\end{align}}
\end{corollary}

\begin{IEEEproof}
See Appendix~\ref{app:MSE_blm_sls}.
\end{IEEEproof} \vspace{1mm}

\noindent The result of Corollary~\ref{cor:MSE_blm_sls} states that, under the above assumptions, the BLM estimator can be implemented in a way that avoids the inversion of and the multiplication with the $M \tau$-dimensional matrix $\Sigmab_{\rmp}$, where $M \tau$ can be quite large at mmWave and sub-THz frequencies. In particular, $\P$ diagonalizes $\Sigmab_{\rmp}$ when $\P \P^{\herm}$ is circulant, which greatly simplifies the structure of \eqref{eq:H_hat_blm}. Lastly, we point out that, in the case of correlated channels, Corollary~\ref{cor:MSE_blm_sls} does not generally hold as the channel covariance matrix is embedded into $\Sigmab_{\rmp}$ and the latter is not diagonalized by $\P$.

Let us move our focus to the class of scaled LS estimators, such as the one in \eqref{eq:H_hat_sls}; recall that, unlike the simplified expression of the BLM estimator in \eqref{eq:H_hat_blm_sls}, scaled LS estimators are characterized by a common scaling factor for all the UEs. In this regard, we first derive the closed-form expression of the MSE of the channel estimation for an arbitrary scaling factor.

\begin{theorem} \label{thm:MSE_sls} \rm{
Suppose that the scaled LS estimator in \eqref{eq:H_hat_sls} is used with arbitrary $\Psi$. Then, the MSE of the channel estimation is given by
\begin{align}
\label{eq:MSE_sls_0} \mse_{\sls} & \triangleq \frac{1}{M K} \Exp \big[ \| \hat{\H}_{\sls} - \H \|_{\rmF}^{2} \big] \\
\label{eq:MSE_sls_gen} & = 1 + (\rho K + 1) \Psi (\tau + \Delta) - 2 \sqrt{\frac{2}{\pi} \rho \Psi} \tau
\end{align}
where we have defined
\begin{align} \label{eq:Delta}
\Delta \triangleq \frac{1}{K} \sum_{k=1}^{K} \delta_{k}
\end{align}
with $\delta_{k}$ defined in \eqref{eq:delta_k}.}
\end{theorem}

\begin{IEEEproof}
The expression in \eqref{eq:MSE_sls_gen} is obtained from the proof of Corollary~\ref{cor:MSE_blm_sls} by replacing the UE-specific scaling factors $\{\Psi_{1}, \ldots, \Psi_{K}\}$ in \eqref{eq:MSEp_blm_sls2} with the common scaling factor $\Psi$.
\end{IEEEproof} \vspace{1mm}

\noindent In particular, when $\Psi$ defined in \eqref{eq:Psi} is used, \eqref{eq:MSE_sls_gen} becomes
\begin{align}
\nonumber \mse_{\sls} & = 1 - \frac{2}{\pi} \rho \bigg( \frac{2}{\pi} \rho (\tau - K) + \rho K + 1 \bigg)^{-2} \\
\label{eq:MSE_sls} & \phantom{=} \ \times \bigg( \frac{4}{\pi} \rho \tau (\tau - K) + (\rho K + 1) (\tau - \Delta) \bigg).
\end{align}
Note that, when $\tau = K$, \eqref{eq:MSE_sls} recovers the expression of the MSE of the BLM estimator in \eqref{eq:MSE_blm_tau=K}.

Now, if we consider the scaling factor $\Psi$ as a tuning parameter, we can minimize $\mse_{\sls}$ in \eqref{eq:MSE_sls_gen} by optimizing over $\Psi$. As a result, we obtain the optimal estimator within the class of scaled LS estimators.

\begin{corollary} \label{cor:MSE_sls_prime} \rm{
Suppose that the scaled LS estimator
\begin{align} \label{eq:H_hat_sls_prime}
\hat{\H}_{\sls}^{\prime} \triangleq \sqrt{\Psi^{\prime}} \R_{\rmp} \P \in \Compl^{M \times K}
\end{align}
is used, where $\Psi^{\prime}$ is obtained by minimizing \eqref{eq:MSE_sls_gen} with respect to $\Psi$ and is defined as
\begin{align} \label{eq:Psi_prime}
\Psi^{\prime} & \triangleq \frac{2}{\pi} \frac{\rho \tau^{2}}{(\rho K + 1)^2 (\tau + \Delta)^{2}}
\end{align}
with $\Delta$ defined in \eqref{eq:Delta}. Then, the MSE of the channel estimation is given by
\begin{align}
\mse_{\sls}^{\prime} & \triangleq \frac{1}{M K} \Exp \big[ \| \hat{\H}_{\sls}^{\prime} - \H \|_{\rmF}^{2} \big] \\
\label{eq:MSE_sls_prime} & = 1 - \frac{2}{\pi} \frac{\rho \tau^{2}}{(\rho K + 1) (\tau + \Delta)} \\
& \leq \eqref{eq:MSE_sls}.
\end{align}}
\end{corollary}

\begin{IEEEproof}
Since \eqref{eq:MSE_sls_gen} is a convex function of $\Psi$, $\Psi^{\prime}$ in \eqref{eq:Psi_prime} can be obtained by setting ${\frac{\diff}{\diff \Psi} \eqref{eq:MSE_sls_gen} = 0}$. Then, replacing $\Psi$ with $\Psi^{\prime}$ in \eqref{eq:MSE_sls_gen} yields the expression in \eqref{eq:MSE_sls_prime}.
\end{IEEEproof} \vspace{1mm}

\noindent Note that $\Psi^{\prime}$ in \eqref{eq:Psi_prime} implicitly depends on the channel distribution as does $\Psi$ in \eqref{eq:Psi}. The result of Corollary~\ref{cor:MSE_sls_prime} shows that the optimal scaled LS estimator (in terms of MSE of the channel estimation) is not the one that simply ignores the temporal correlation of the quantization distortion from the BLM estimator (see \eqref{eq:H_hat_sls}--\eqref{eq:Psi}); instead, a simple optimization over the scaling factor can significantly improve the channel estimation accuracy. When $\tau = K$, the optimal scaled LS estimator in \eqref{eq:H_hat_sls_prime} coincides with the estimator in \eqref{eq:H_hat_sls}, and, in turn, with the BLM estimator in \eqref{eq:H_hat_blm}: in fact, $\tau = K$ implies that $\Psi^{\prime}$ in \eqref{eq:Psi_prime} reduces to $\Psi$ in \eqref{eq:Psi} and \eqref{eq:MSE_sls_prime} recovers the expression in \eqref{eq:MSE_blm_tau=K}. On the other hand, the estimator in \eqref{eq:H_hat_sls_prime} shall be always preferred to the estimator in \eqref{eq:H_hat_sls} when $\tau > K$ and the performance gap between the two widens with $\tau - K$. The improved performance of \eqref{eq:H_hat_sls_prime} over \eqref{eq:H_hat_sls} is also suggested by the resemblance between $\mse_{\blm}$ in \eqref{eq:MSE_blm} and $\mse_{\sls}^{\prime}$ in \eqref{eq:MSE_sls_prime}, where the latter can be obtained from the former by replacing $\delta_{k}$ with $\Delta$, $\forall k$. Remarkably, in Section~\ref{sec:PA_EST_num}, we show that the optimal scaled LS estimator in \eqref{eq:H_hat_sls_prime} entails a negligible performance loss with respect to the BLM estimator.

It is of particular interest to study the asymptotic behavior of the MSE of the channel estimation at high SNR.

\begin{corollary} \label{cor:MSE-lim_rho} \rm{
From Theorems~\ref{thm:MSE_blm} and~\ref{thm:MSE_sls} and from Corollary~\ref{cor:MSE_sls_prime}, in the limit of $\rho \to \infty$, we have
\begin{align}
\label{eq:MSE_blm-lim_rho} \lim_{\rho \to \infty} \mse_{\blm} & = 1 - \frac{2}{\pi} \frac{\tau^{2}}{K^{2}} \sum_{k=1}^{K} \frac{1}{\tau + \bar{\delta}_{k}}, \\
\nonumber \lim_{\rho \to \infty} \mse_{\sls} & = 1 - \frac{2}{\pi} \bigg( \frac{2}{\pi} (\tau - K) + K \bigg)^{-2} \\
\label{eq:MSE_sls-lim_rho} & \phantom{=} \ \times \bigg( \frac{4}{\pi} \tau (\tau - K) + K (\tau - \bar{\Delta}) \bigg), \\
\label{eq:MSE_sls_prime-lim_rho} \lim_{\rho \to \infty} \mse_{\sls}^{\prime} & = 1 - \frac{2}{\pi} \frac{\tau^{2}}{K (\tau + \bar{\Delta})}
\end{align}
where we have defined
\begin{align}
\nonumber \bar{\delta}_{k} & \triangleq \sum_{u \neq v} \bigg( \Re[P_{u,k}^{*} P_{v,k}] \Omega \bigg( \frac{\sum_{i=1}^{K} \Re[P_{u,i} P_{v,i}^{*}]}{K} \bigg) \\
\label{eq:delta_bar_k} & \phantom{=} \ - \Im[P_{u,k}^{*} P_{v,k}] \Omega \bigg( \frac{\sum_{i=1}^{K} \Im[P_{u,i} P_{v,i}^{*}]}{K} \bigg) \bigg)
\end{align}
and
\begin{align} \label{eq:Delta_bar}
\bar{\Delta} \triangleq \frac{1}{K} \sum_{k=1}^{K} \bar{\delta}_{k}.
\end{align}}
\end{corollary}

\noindent The results of Corollary~\ref{cor:MSE-lim_rho} show that arbitrarily increasing the transmit SNR is detrimental for the performance of the channel estimation since the right amount of noise is necessary to recover the difference in amplitude between channel entries (see Appendix~\ref{app:ch_est}). This is in sheer contrast with the case of infinite-resolution ADCs, where boosting $\rho$ produces the same beneficial noise-averaging effect as increasing $\tau$. In the next section, we also discuss the asymptotic behavior at low SNR.

\subsection{Tractable Upper Bounds} \label{sec:PA_EST_ub}

The MSE expressions derived so far depend on the specific pilot choice through the parameters $\{\delta_{1}, \ldots, \delta_{K}\}$ in \eqref{eq:delta_k} or $\Delta$ in \eqref{eq:Delta}. To gain more practical insights, we now consider the single-UE case (i.e., $K = 1$) and derive tractable upper bounds that are independent of the pilot choice. We begin by pointing out that, when $K = 1$, we have $\delta_{k} = \Delta$ and, thus, $\mse_{\blm}$ in \eqref{eq:MSE_blm} is equal to $\mse_{\sls}^{\prime}$ in \eqref{eq:MSE_sls_prime}. Let $\p \triangleq (p_{u}) \in \Compl^{\tau \times 1}$ denote the pilot used by the UE. In this setting, $\Delta$ in \eqref{eq:Delta} can be simplified as
\begin{align}
\nonumber \Delta & = \sum_{u \neq v} \bigg( \Re[p_{u}^{*} p_{v}] \Omega \bigg( \frac{\rho \Re[p_{u} p_{v}^{*}]}{\rho + 1} \bigg) \\
\label{eq:delta} & \phantom{=} \ - \Im[p_{u}^{*} p_{v}] \Omega \bigg( \frac{\rho \Im[p_{u} p_{v}^{*}]}{\rho + 1} \bigg) \bigg) \\
\label{eq:Delta-p=1} & \leq \tau (\tau - 1) \Omega \bigg( \frac{\rho}{\rho + 1} \bigg)
\end{align}
where the upper bound in \eqref{eq:Delta-p=1} is obtained by fixing $\p$ such that $p_{u} \in \{\pm \beta, \pm j \, \beta\}$, $\forall u$, with $\beta \in \Compl$ and $|\beta|^{2} = 1$; in the rest of the paper, when referring to this case, we will simply use $\p = \1_{\tau}$. Indeed, for a given pilot length $\tau$, such a structure of $\p$ represents the worst possible pilot choice since it maximizes the MSE of the channel estimation in \eqref{eq:MSE_blm} and \eqref{eq:MSE_sls}. As detailed in Appendix~\ref{app:ch_est}, this effect is particularly detrimental at high SNR and, in the limit of $\rho \to \infty$, each channel entry is reduced to a scaled symbol of the QPSK constellation regardless of the value of $\tau$.

\begin{figure*}[t!]
\centering
\begin{subfigure}{0.49\textwidth}
\centering
\includegraphics[scale=1]{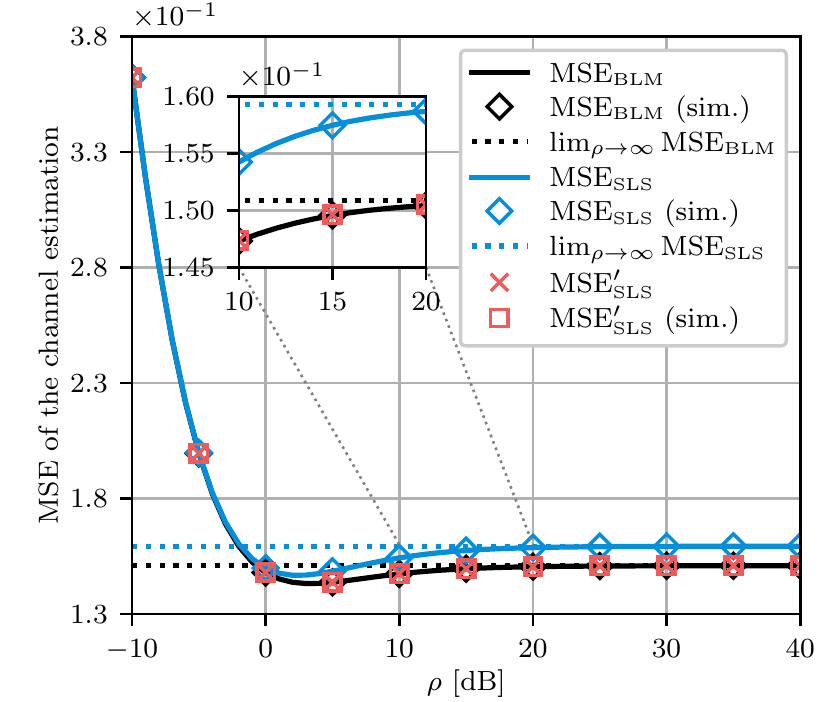}
\vspace{-0.5mm} \caption{$K = 4$, $\tau = 32$.}
\end{subfigure} \hfill
\begin{subfigure}{0.49\textwidth}
\centering
\includegraphics[scale=1]{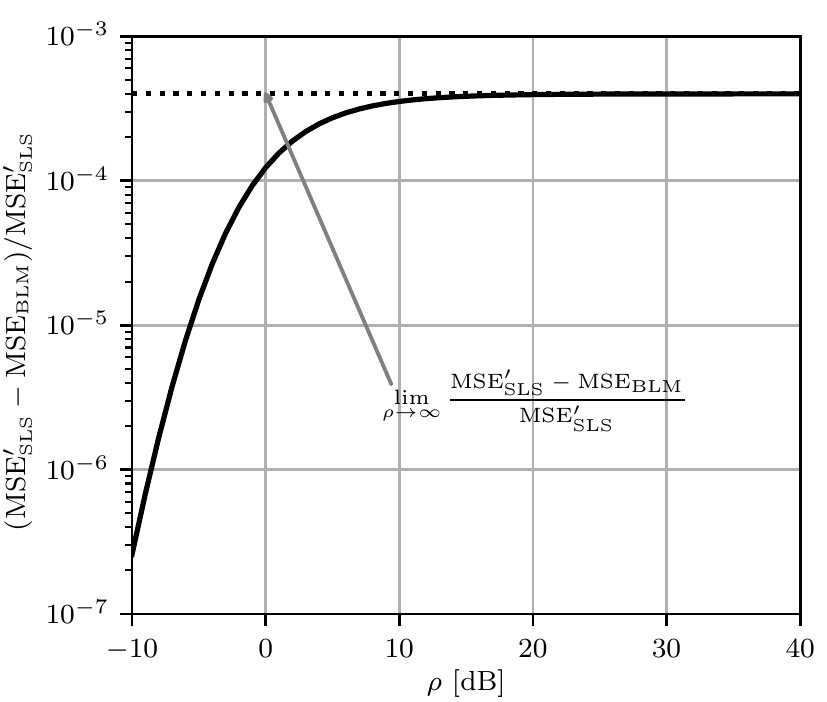}
\vspace{-0.5mm} \caption{$K = 4$, $\tau = 32$.}
\end{subfigure} \\
\begin{subfigure}{0.49\textwidth}
\centering
\includegraphics[scale=1]{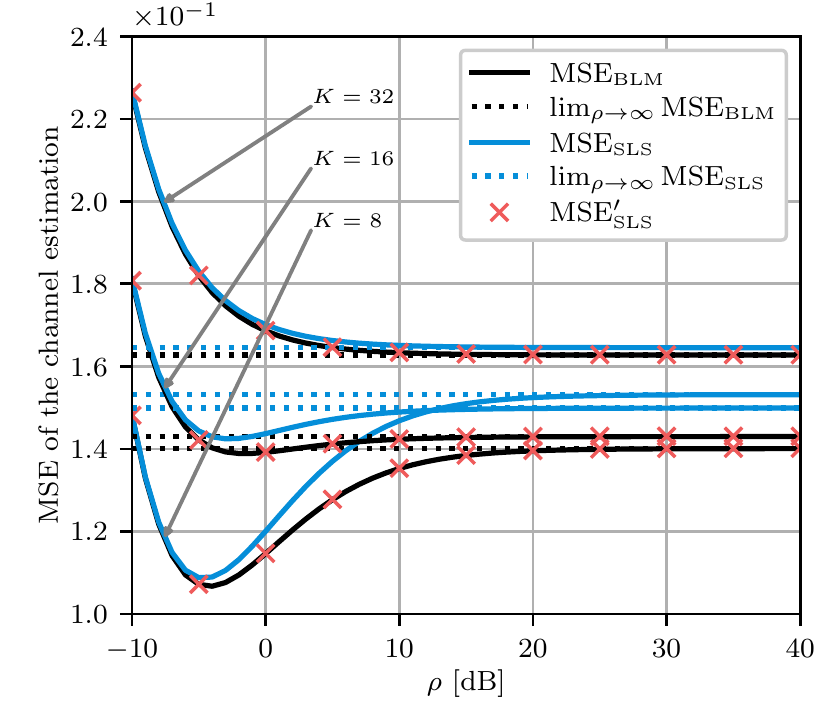}
\vspace{-0.5mm} \caption{$K \in \{8, 16, 32\}$, $\tau = 128$.}
\end{subfigure} \hfill
\begin{subfigure}{0.49\textwidth}
\centering
\includegraphics[scale=1]{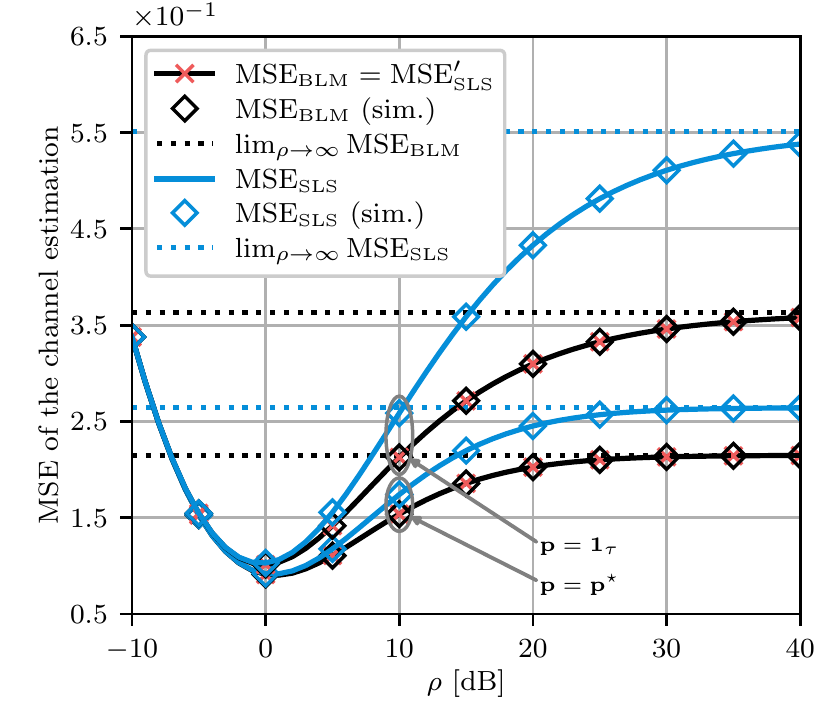}
\vspace{-0.5mm} \caption{$K = 1$, $\tau = 32$.}
\end{subfigure}
\caption{MSE of the channel estimation against the transmit SNR.} \label{fig:MSE_VS_rho} \vspace{-5.5mm}
\end{figure*}

Hence, plugging \eqref{eq:Delta-p=1} into \eqref{eq:MSE_blm} and \eqref{eq:MSE_blm-lim_rho} yields
\begin{align} \label{eq:MSE_blm-p=1}
\mse_{\blm} & = 1 - \frac{2}{\pi} \frac{\rho \tau}{(\rho + 1) \big( 1 + (\tau - 1) \Omega \big( \frac{\rho}{\rho + 1} \big) \big)}
\end{align}
and
\begin{align} \label{eq:MSE_blm-lim_rho-p=1}
\lim_{\rho \to \infty} \mse_{\blm} & = 1 - \frac{2}{\pi}
\end{align}
respectively. In addition, considering \eqref{eq:MSE_blm-p=1} in the limit of $\tau \to \infty$, we have
\begin{align} \label{eq:MSE_blm-lim_tau-p=1}
\lim_{\tau \to \infty} \mse_{\blm} & = 1 - \frac{2}{\pi} \frac{\rho}{(\rho + 1) \Omega \big( \frac{\rho}{\rho + 1} \big)}.
\end{align}
Likewise, plugging \eqref{eq:Delta-p=1} into \eqref{eq:MSE_sls} and \eqref{eq:MSE_sls-lim_rho} yields
\begin{align}
\nonumber \mse_{\sls} & = 1 - \frac{2}{\pi} \rho \tau \bigg( \frac{2}{\pi} \rho (\tau - 1) + \rho + 1 \bigg)^{-2} \bigg( \frac{4}{\pi} \rho (\tau - 1) \\
\label{eq:MSE_sls-p=1} & \phantom{=} \ + (\rho + 1) \bigg( 1 - (\tau - 1) \Omega \bigg( \frac{\rho}{\rho + 1} \bigg) \bigg) \bigg)
\end{align}
and
\begin{align} \label{eq:MSE_sls-lim_rho-p=1}
\lim_{\rho \to \infty} \mse_{\sls} & = 1 \! - \! \frac{2}{\pi} \tau \bigg( \! \frac{2}{\pi} (\tau \! - \! 1) \! + \! 1 \! \bigg)^{-2} \bigg( \! \frac{4}{\pi} (\tau \! - \! 1) \! - \! \tau \! + \! 2 \! \bigg)
\end{align}
respectively. Moreover, considering \eqref{eq:MSE_sls-p=1} in the limit of $\tau \to \infty$, we have
\begin{align} \label{eq:MSE_sls-lim_tau-p=1}
\lim_{\tau \to \infty} \mse_{\sls} & = \frac{\pi}{2} \frac{\rho + 1}{\rho} \Omega \bigg( \frac{\rho}{\rho + 1} \bigg) - 1.
\end{align}
Some comments are in order. First, when $\tau = 1$, both \eqref{eq:MSE_blm-p=1} and \eqref{eq:MSE_sls-p=1} recover the expression in \eqref{eq:MSE_blm_tau=K} with $K=1$. Second, \eqref{eq:MSE_blm-lim_rho-p=1} does not depend on $\tau$ since, in the absence of noise, estimating the channel repeatedly over the same pilot symbol does not bring any benefit. Third, it can be demonstrated that \eqref{eq:MSE_blm-p=1} and \eqref{eq:MSE_sls-p=1} are quasiconvex functions of $\rho$ and, as such, they have a unique minimum. This defines a clear SNR trade-off, according to which operating at the right noise level enhances the channel estimation accuracy. In particular, as discussed in Appendix~\ref{app:ch_est}, we have that:
\begin{itemize}
\item[$\bullet$] At low SNR, the channel estimates are corrupted by the strong noise;
\item[$\bullet$] At high SNR, the difference in amplitude between channel entries cannot be recovered.
\end{itemize}
In general, when $\tau > 1$, the value of $\rho$ that minimizes \eqref{eq:MSE_blm-p=1}, denoted by $\rho^{\star}$, satisfies
\begin{align} \label{eq:rho_star_est}
\frac{2}{\pi} \frac{\rho^{\star}}{\sqrt{1 + 2 \rho^{\star}}} - \Omega \bigg( \frac{\rho^{\star}}{\rho^{\star} + 1} \bigg) = \frac{1}{\tau - 1}.
\end{align}
Since the left-hand side of \eqref{eq:rho_star_est} monotonically increases with the transmit SNR, $\rho^{\star}$ decreases as the left-hand side of \eqref{eq:rho_star_est} decreases, i.e., as $\tau$ increases. This means that using longer pilots allows to operate at lower SNR as the noise can be averaged out more efficiently. This interdependence between $\rho$ and $\tau$ can be also observed from \eqref{eq:MSE_blm-lim_tau-p=1}: in the limit of $\tau \to \infty$, since $\lim_{w \to 0} \frac{w}{\arcsin(w)} = 1$, it follows that $\mse^{\prime} \to 0$ as $\rho \to 0$. Lastly, it is shown in Section~\ref{sec:PA_EST_num} that the upper bounds obtained by fixing $\p = \1_{\tau}$ are remarkably tight at low SNR and up to the region around the optimal value of $\rho$. Therefore, the above observations also apply to the general~case.

\begin{figure*}[t!]
\centering
\begin{subfigure}{0.49\textwidth}
\centering
\includegraphics[scale=1]{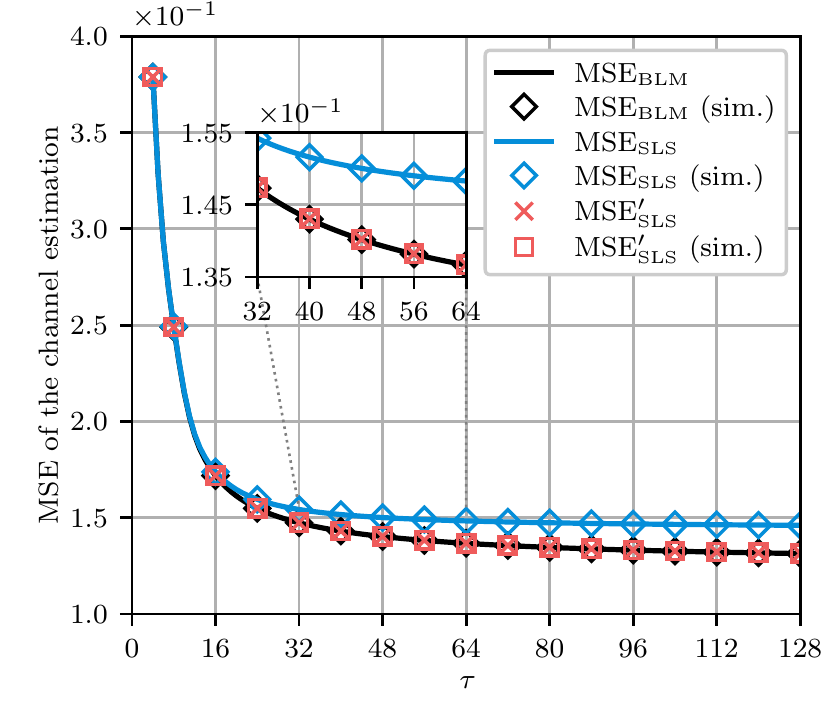}
\vspace{-0.5mm} \caption{$K = 4$, $\rho = 10$~dB.}
\end{subfigure} \hfill
\begin{subfigure}{0.49\textwidth}
\centering
\includegraphics[scale=1]{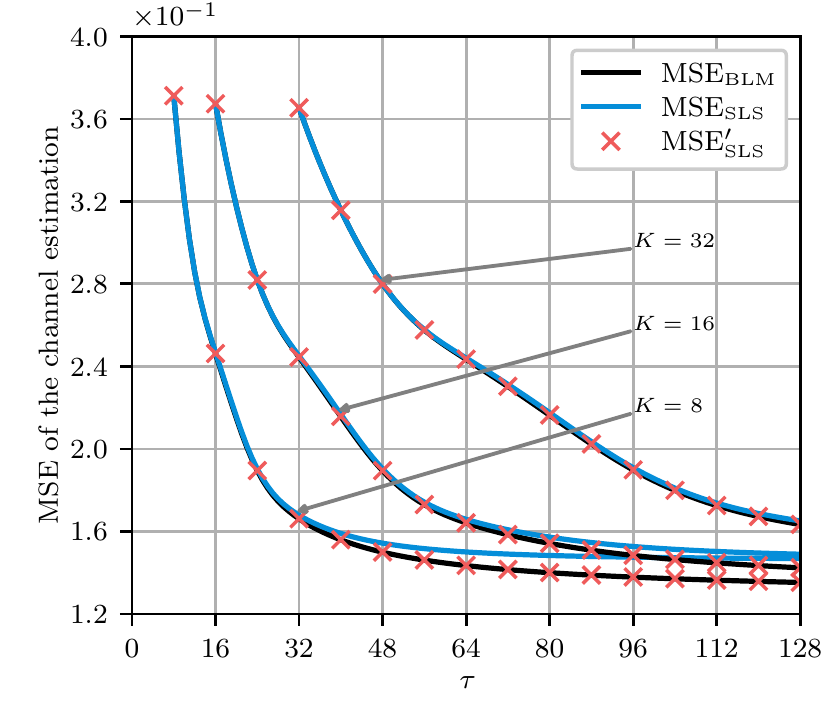}
\vspace{-0.5mm} \caption{$K \in \{8, 16, 32\}$, $\rho = 10$~dB.}
\end{subfigure} \\
\begin{subfigure}{0.49\textwidth}
\centering
\includegraphics[scale=1]{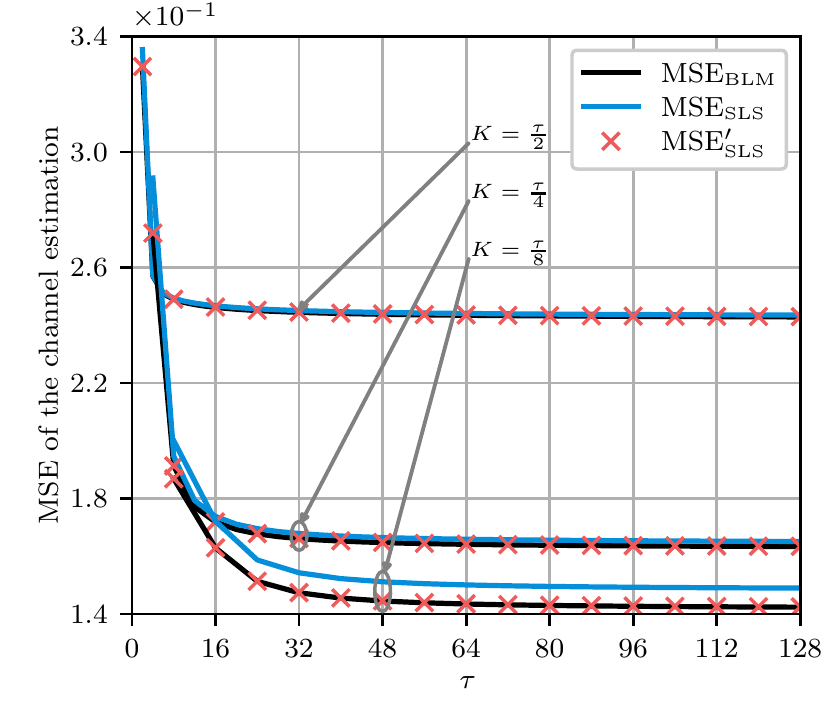}
\vspace{-0.5mm} \caption{$K \in \big\{ \frac{\tau}{8}, \frac{\tau}{4}, \frac{\tau}{2} \big\}$, $\rho = 10$~dB.}
\end{subfigure} \hfill
\begin{subfigure}{0.49\textwidth}
\centering \vspace{0.25mm}
\includegraphics[scale=1]{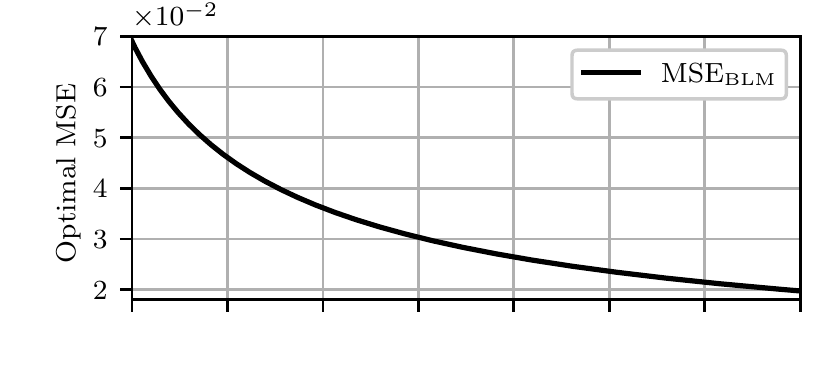} \\ \vspace{-8.25mm}
\includegraphics[scale=1]{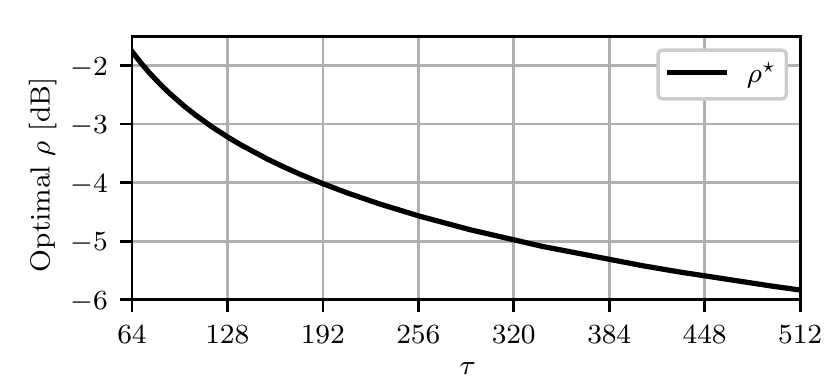}
\vspace{-0.5mm} \caption{$K = 1$, $\p = \1_{\tau}$.}
\end{subfigure}
\caption{MSE of the channel estimation against the pilot length.} \label{fig:MSE_VS_tau} \vspace{-5.5mm}
\end{figure*}

\subsection{Numerical Results and Discussion} \label{sec:PA_EST_num}

We now focus on the performance evaluation of the channel estimation with 1-bit ADCs with respect to the different parameters based on the analytical results presented in Sections~\ref{sec:PA_EST_main} and~\ref{sec:PA_EST_ub}. In this regard, when $K > 1$, we choose a pilot matrix $\P$ composed of the first $K$ columns of the $\tau$-dimensional DFT matrix. On the other hand, when $K = 1$, we use the pilots $\p = \p^{\star}$ and $\p = \1_{\tau}$, where
\begin{align} \label{eq:p^star}
\p^{\star} \triangleq [1, e^{-j \, \frac{\pi}{2 \tau}}, e^{-j \, 2 \frac{\pi}{2 \tau}}, \ldots, e^{-j \, (\tau-1) \frac{\pi}{2 \tau}}]^{\tran} \in \Compl^{\tau \times 1}
\end{align}
denotes the vector whose entries are equispaced on the first quadrant of the unit circle: these represent the best and the worst possible pilot choices, respectively (see Section~\ref{sec:PA_EST_ub} and Appendix~\ref{app:ch_est} for more details on the pilot choice). We thus consider the expressions of $\mse_{\blm}$ derived in \eqref{eq:MSE_blm}, resulting from the BLM estimator in \eqref{eq:H_hat_blm} (see \cite{Li17}), $\mse_{\sls}$ derived in \eqref{eq:MSE_sls}, resulting from the scaled LS estimator in \eqref{eq:H_hat_sls} (see \cite{Mol17}), and $\mse_{\sls}^{\prime}$ derived in \eqref{eq:MSE_sls_prime}, resulting from the optimal scaled LS estimator in \eqref{eq:H_hat_sls_prime}. These are compared with Monte Carlo simulations with $10^{6}$ independent channel realizations. For the latter, we fix $M = K$ to generate the channel matrices, although the value of $M$ does not affect the analytical and numerical results in any way.

Fig.~\ref{fig:MSE_VS_rho} illustrates the MSE of the channel estimation against the transmit SNR $\rho$, also including the asymptotic MSE expressions in \eqref{eq:MSE_blm-lim_rho}--\eqref{eq:MSE_sls_prime-lim_rho}. Fig.~\ref{fig:MSE_VS_rho}(a) considers $K = 4$ and $\tau = 32$ and shows that $\mse_{\blm}$ is $5.3\%$ lower than $\mse_{\sls}$ at high SNR. Remarkably, the performance loss associated with $\mse_{\sls}^{\prime}$ with respect to $\mse_{\blm}$ is negligible and can only be noticed by examining the relative MSE difference in Fig.~\ref{fig:MSE_VS_rho}(b), which reaches its maximum of about $4 \times 10^{-4}$ in the limit of $\rho \to \infty$. Hence, in the case of i.i.d. Rayleigh fading channels among the UEs, the scaled LS estimator with a common optimized scaling factor for all the UEs essentially achieves the same accuracy as the BLM estimator. Lastly, we highlight the SNR trade-off described in Sections~\ref{sec:PA_EST_main} and~\ref{sec:PA_EST_ub} as well as in Appendix~\ref{app:ch_est}, whereby the MSE of the channel estimation exhibits a valley at about $\rho = 3$~dB. Fig.~\ref{fig:MSE_VS_rho}(c) considers $\tau = 128$ and different values of $K$. Here, the gap between $\mse_{\blm}$ and $\mse_{\sls}$ widens as $\tau - K$ increases, reaching $8.5\%$ for $K = 8$ at high SNR. In this respect, at high SNR, $\mse_{\sls}$ for $K = 8$ surpasses its counterpart for $K = 16$: in fact, \eqref{eq:MSE_blm-lim_rho}--\eqref{eq:MSE_sls_prime-lim_rho} are not monotonically increasing with $K$ due to the fact that $\bar{\delta}_{k}$ in \eqref{eq:delta_bar_k} is a decreasing function of $K$. Moreover, the SNR trade-off appears more evident for small values of $K$. Fig.~\ref{fig:MSE_VS_rho}(d) considers the single-UE case, showing that the upper bounds in \eqref{eq:MSE_blm-p=1} and \eqref{eq:MSE_sls-p=1} obtained by fixing $\p = \1_{\tau}$ are remarkably tight at low SNR and up to the region around the optimal transmit SNR. Note that the optimal value of $\rho$ with $\p = \1_{\tau}$ satisfies the condition in \eqref{eq:rho_star_est} and gives an accurate approximation of the optimal~value~of~$\rho$~with~$\p = \p^{\star}$.

Fig.~\ref{fig:MSE_VS_tau} plots the MSE of the channel estimation against the pilot length $\tau$. The transmit SNR is fixed to $\rho = 10$~dB in Fig.~\ref{fig:MSE_VS_tau}(a)--(c), whereas Fig.~\ref{fig:MSE_VS_tau}(d) considers the optimized transmit SNR (we recall that $\rho$ should be reduced as $\tau$ increases to enhance the channel estimation accuracy). Fig.~\ref{fig:MSE_VS_tau}(a) considers $K = 4$ and shows that $\mse_{\blm}$ is $10\%$ lower than $\mse_{\sls}$ at $\tau = 128$. Furthermore, as in Fig.~\ref{fig:MSE_VS_rho}, $\mse_{\sls}^{\prime}$ closely matches $\mse_{\blm}$ for any value of $\tau$, which means that the optimal scaled LS estimator essentially achieves the same accuracy as the BLM estimator. Fig.~\ref{fig:MSE_VS_tau}(b) considers different values of $K$, showing that the gap between $\mse_{\blm}$ and $\mse_{\sls}$ widens as $\tau - K$ increases and reaches $7.8\%$ for $K = 8$ and $\tau = 128$. Fig.~\ref{fig:MSE_VS_tau}(c) examines the case where the number of UEs grows together with the pilot length. In this setting, the gap between $\mse_{\blm}$ and $\mse_{\sls}$ is roughly constant and increases with the ratio $\frac{\tau}{K}$, reaching about $5\%$ for $\frac{\tau}{K} = 8$. Lastly, Fig.~\ref{fig:MSE_VS_tau}(d) considers the single-UE case and the upper bound on $\mse_{\blm}$ in \eqref{eq:MSE_blm-p=1} obtained by fixing $\p = \1_{\tau}$, which is optimized over the transmit SNR for each $\tau$. As discussed in Section~\ref{sec:PA_EST_ub}, the optimal value of $\rho$ satisfies the condition in \eqref{eq:rho_star_est} and decreases as $\tau$ increases.

\section{Data Detection with 1-bit ADCs and MRC} \label{sec:PA_DEC}

In this section, we are interested in characterizing the performance of the data detection with respect to the different parameters when 1-bit ADCs are adopted at each BS antenna (see Section~\ref{sec:SM_dec}). In this regard, we consider the scenario where the BS uses the BLM estimator in \eqref{eq:H_hat_blm} in the channel estimation phase and the MRC receiver in the data detection phase (see also \cite{Li17}). Assuming that $\P \P^{\herm}$ is circulant and building on Corollary~\ref{cor:MSE_blm_sls}, the combining matrix is given by $\V = \hat{\H}_{\blm}$ and we can write the soft estimate in \eqref{eq:x_hat} as
\begin{align}
\hat{\x} & = \Psib^{\frac{1}{2}} \P^{\herm} \R_{\rmp}^{\herm} \r \\
\nonumber & = \frac{\rho K + 1}{2} \Psib^{\frac{1}{2}} \P^{\herm} \Big( \sgn \big( \Re[\sqrt{\rho} \H \P^{\herm} + \Z_{\rmp}] \big) \\
\nonumber & \phantom{=} \ + j \, \sgn \big( \Im[\sqrt{\rho} \H \P^{\herm} + \Z_{\rmp}] \big) \Big)^{\herm} \\
\label{eq:x_hat_mrc} & \phantom{=} \ \times \Big( \sgn \big( \Re[\sqrt{\rho} \H \x + \z] \big) + j \, \sgn \big( \Im[\sqrt{\rho} \H \x + \z] \big) \Big).
\end{align}
In the following, we focus on the single-UE case (i.e., $K = 1$) and characterize the statistical properties of the estimated symbols. We recall that the BLM estimator is equivalent to the optimal scaled LS estimator in \eqref{eq:H_hat_sls_prime} when $K = 1$, as detailed in Section~\ref{sec:PA_EST_ub}. Note that, in a multi-UE massive MIMO context with infinite-resolution ADCs, MRC asymptotically becomes the optimal receive strategy as the number of BS antennas increases. However, when the MRC receiver results from the quantized channel estimation, it cannot be perfectly aligned with the channel matrix, resulting in residual multi-UE interference. Hence, the following analysis of the single-UE case does not consider this interference; nonetheless, this can be straightforwardly included at the expense of more involved and less insightful expressions, which is left for future work.

\subsection{Expected Value and Variance of the Estimated Symbols} \label{sec:PA_DEC_main}

Let $x \in \setS$ be the transmit symbol of the UE, where $\setS \triangleq \{s_{1}, \ldots, s_{L}\}$ denotes the set of transmit symbols, with $s_{\ell} \in \Compl$, $\forall \ell$; for instance, $\setS$ may correspond to the QPSK or 16-QAM constellation. To facilitate the data detection process at the BS, for each transmit symbol $s_{\ell} \in \setS$, we are interested in deriving the closed-form expression of the expected value of the resulting estimated symbol $\hat{s}_{\ell}$.

\begin{theorem} \label{thm:s_mean} \rm{
Assuming $K = 1$ and MRC, for each transmit symbol $s_{\ell} \in \setS$, the expected value of the resulting estimated symbol $\hat{s}_{\ell}$, denoted by $\El \triangleq \Exp [\hat{s}_{\ell}]$, is given by
\begin{align}
\nonumber \El & = \sqrt{\frac{2}{\pi} \rho} M \frac{\tau}{\tau + \Delta} \sum_{u=1}^{\tau} p_{u}^{*} \bigg( \Omega \bigg( \frac{\rho \Re[p_{u} s_{\ell}]}{\sqrt{(\rho + 1)(\rho |s_{\ell}|^{2} + 1)}} \bigg) \\
\label{eq:s_mean} & \phantom{=} \ + j \, \Omega \bigg( \frac{\rho \Im[p_{u} s_{\ell}]}{\sqrt{(\rho + 1)(\rho |s_{\ell}|^{2} + 1)}} \bigg) \bigg)
\end{align}
with $\Delta$ given in \eqref{eq:delta}.}
\end{theorem}

\begin{IEEEproof}
See Appendix~\ref{app:s_mean}.
\end{IEEEproof} \vspace{1mm}

\noindent The result of Theorem~\ref{thm:s_mean} can be exploited to efficiently implement MDD. In this context, each estimated symbol $\hat{x}$ resulting from transmitting $x \in \setS$ can be readily mapped to one of the expected values $\{\mathsf{E}_{1}, \ldots, \mathsf{E}_{L}\}$, which are derived as in \eqref{eq:s_mean} without any prior Monte Carlo computation, according to the minimum distance criterion. To further simplify the process and avoid computing the distance between $\hat{x}$ and each $\El$, one can construct the Voronoi tessellation based on $\{\mathsf{E}_{1}, \ldots, \mathsf{E}_{L}\}$ (and, possibly, other available information) obtaining well-defined detection regions.

Now, for each transmit symbol $s_{\ell} \in \setS$, we derive the closed-form expression of the variance of the resulting estimated symbol $\hat{s}_{\ell}$.

\begin{theorem} \label{thm:s_var} \rm{
Assuming $K = 1$ and MRC, for each transmit symbol $s_{\ell} \in \setS$, the variance of the resulting estimated symbol $\hat{s}_{\ell}$, denoted by $\Vl \triangleq \Var [\hat{s}_{\ell}]$, is given by
\begin{align}
\label{eq:s_var} \Vl & = \frac{2}{\pi} \rho M \frac{\tau^{2}}{\tau + \Delta} - \frac{1}{M} |\El|^{2}
\end{align}
with $\Delta$ given in \eqref{eq:delta} and where $\El$ is derived in closed form in \eqref{eq:s_mean}.}
\end{theorem}

\begin{IEEEproof}
See Appendix~\ref{app:s_var}.
\end{IEEEproof} \vspace{1mm}

\noindent The result of Theorem~\ref{thm:s_var} quantifies the dispersion of the estimated symbols about their expected value, which results from the 1-bit quantization applied to both the channel estimation (through the MRC receiver) and the uplink data transmission. This dispersion is not isotropic and assumes different shapes for different transmit symbols, as illustrated in Section~\ref{sec:PA_DEC_num} (see also \cite{Abd21}). Some additional comments are in order. First, $\Vl$ reduces as $|s_{\ell}|$ increases due to the negative term on the right-hand side of \eqref{eq:s_var}: this is somewhat intuitive since the transmit symbols that lie further from the origin are less subject to noise. Second, although $\Vl$ increases linearly with the number of BS antennas $M$, the normalized variance $\frac{\Vl}{|\El|^{2}}$ (which expresses the relative dispersion of the estimated symbols about their expected value) is inversely proportional to $M$. Third, the combined results of Theorems~\ref{thm:s_mean} and~\ref{thm:s_var} can be exploited to design the set of transmit symbols $\setS$ by jointly minimizing the relative dispersion and the overlap between different symbols after the estimation, which is left for future work. Lastly, in the context of MDD via Voronoi tessellation described above, one can utilize the variance derived as in \eqref{eq:s_var} to further refine the detection regions \cite{Atz21}.

It is of particular interest to study the asymptotic behavior of the expected value and variance of the estimated symbols at high SNR.

\begin{corollary} \label{cor:decoding-lim_rho} \rm{
From Theorems~\ref{thm:s_mean} and~\ref{thm:s_var}, in the limit of $\rho \to \infty$, we have
\begin{align}
\nonumber \lim_{\rho \to \infty} \frac{\El}{\sqrt{\rho}} & = \sqrt{\frac{2}{\pi}} M \frac{\tau}{\tau + \bar{\Delta}} \sum_{u=1}^{\tau} p_{u}^{*} \bigg( \Omega \bigg( \frac{\Re[p_{u} s_{\ell}]}{|s_{\ell}|} \bigg) \\
\label{eq:s_mean-lim_rho} & \phantom{=} \ + j \, \Omega \bigg( \frac{\Im[p_{u} s_{\ell}]}{|s_{\ell}|} \bigg) \bigg)
\end{align}
and
\begin{align}
\label{eq:s_var-lim_rho} \lim_{\rho \to \infty} \frac{\Vl}{\rho} & = \frac{2}{\pi} M \frac{\tau^{2}}{\tau + \bar{\Delta}} - \frac{1}{M} \lim_{\rho \to \infty} \frac{|\El|^{2}}{\rho}
\end{align}
with $\bar{\Delta}$ defined in \eqref{eq:Delta_bar}, which can be simplified for $K = 1$ as
\begin{align}
\bar{\Delta} & = \sum_{u \neq v} \Big( \Re[p_{u}^{*} p_{v}] \Omega \big( \Re[p_{u} p_{v}^{*}] \big) - \Im[p_{u}^{*} p_{v}] \Omega \big( \Im[p_{u} p_{v}^{*}] \big) \Big).
\end{align}}
\end{corollary}

\noindent Corollary~\ref{cor:decoding-lim_rho} formalizes a behavior of the estimated symbols that was observed in \cite{Jac17}. From \eqref{eq:s_mean-lim_rho}, it emerges that, at high SNR, all the estimated symbols lie on a circle around the origin and their amplitude no longer conveys any information. As a consequence, the estimated symbols resulting from transmit symbols with the same phase become indistinguishable in terms of their expected value, which depends only on $\frac{\Re[s_{\ell}]}{|s_{\ell}|}$ and $\frac{\Im[s_{\ell}]}{|s_{\ell}|}$. For instance, if $\setS$ corresponds to the 16-QAM constellation as in Section~\ref{sec:PA_DEC_num}, the inner estimated symbols become indistinguishable from the outer estimated symbols with the same phase. Furthermore, according to \eqref{eq:s_var-lim_rho}, these estimated symbols become identical also in terms of variance. In the light of this, blindly minimizing the (normalized) variance of the estimated symbols is not the key to enhancing the system performance. Instead, the variance (which roughly decreases with the transmit SNR) should be minimized alongside the overlap between different symbols after the estimation (which generally increases with the transmit SNR). This determines a clear SNR trade-off, according to which operating at the right noise level enhances the data detection accuracy and thus reduces the SER. In the next section, we also discuss the asymptotic behavior at low SNR.

\subsection{Tractable Upper Bounds} \label{sec:PA_DEC_ub}

As done in Section~\ref{sec:PA_EST_ub} for the MSE of the channel estimation, tractable upper bounds on the normalized variance of the estimated symbols, i.e., that do not depend on the specific pilot choice, can be obtained by fixing $\p = \1_{\tau}$ since such a structure of $\p$ represents the worst possible pilot choice (see Section~\ref{sec:PA_EST_ub} and Appendix~\ref{app:ch_est}). Hence, plugging \eqref{eq:Delta-p=1} into \eqref{eq:s_var} and \eqref{eq:s_var-lim_rho} yields
\begin{align}
\nonumber \hspace{-2.5mm} \frac{\Vl}{|\El|^{2}} & = \frac{1}{M} \frac{1 \! + \! (\tau \! - \! 1) \Omega \big( \frac{\rho}{\rho + 1} \big)}{\tau} \bigg( \! \bigg( \! \Omega \bigg( \! \frac{\rho \Re[s_{\ell}]}{\sqrt{(\rho \! + \! 1)(\rho |s_{\ell}|^{2} \! + \! 1)}} \! \bigg) \! \bigg)^{2} \\
\label{eq:s_var_norm-p=1} & \phantom{=} \ + \bigg( \Omega \bigg( \frac{\rho \Im[s_{\ell}]}{\sqrt{(\rho + 1)(\rho |s_{\ell}|^{2} + 1)}} \bigg) \bigg)^{2} \bigg)^{-1} - \frac{1}{M}
\end{align}
and
\begin{align}
\label{eq:s_var_norm-lim_rho-p=1} \lim_{\rho \to \infty} \frac{\Vl}{|\El|^{2}} & = \frac{1}{M} \bigg( \! \bigg( \! \Omega \bigg( \! \frac{\Re[s_{\ell}]}{|s_{\ell}|} \! \bigg) \! \bigg)^{2} \! + \! \bigg( \! \Omega \bigg( \! \frac{\Im[s_{\ell}]}{|s_{\ell}|} \! \bigg) \! \bigg)^{2} \bigg)^{-1} \! - \! \frac{1}{M}
\end{align}
respectively. In addition, considering \eqref{eq:s_var_norm-p=1} in the limit of $\tau \to \infty$, we have

\newpage

$ $ \vspace{-8mm}

\begin{align}
\nonumber \hspace{-2mm} \lim_{\tau \to \infty} \frac{\Vl}{|\El|^{2}} & = \frac{1}{M} \Omega \bigg( \! \frac{\rho}{\rho \! + \! 1} \! \bigg) \bigg( \! \bigg( \! \Omega \bigg( \! \frac{\rho \Re[s_{\ell}]}{\sqrt{(\rho \! + \! 1)(\rho |s_{\ell}|^{2} \! + \! 1)}} \! \bigg) \! \bigg)^{2} \\
\label{eq:s_var_norm-lim_tau-p=1} & \phantom{=} \ + \! \bigg( \! \Omega \bigg( \! \frac{\rho \Im[s_{\ell}]}{\sqrt{(\rho \! + \! 1)(\rho |s_{\ell}|^{2} \! + \! 1)}} \! \bigg) \! \bigg)^{2} \bigg)^{-1} \! - \! \frac{1}{M}.
\end{align}
Some comments are in order. First, the normalized variance of the estimated symbols can be made arbitrarily close to zero by increasing the number of BS antennas $M$. Second, \eqref{eq:s_var_norm-lim_rho-p=1} does not depend on $\tau$ since, in the absence of noise, estimating the channel repeatedly over the same pilot symbol does not bring any benefit. Third, it can be demonstrated that \eqref{eq:s_var_norm-p=1} is a quasiconvex function of $\rho$ and, as such, it has a unique minimum that defines a further SNR trade-off. It is shown in Section~\ref{sec:PA_DEC_num} that this SNR trade-off, which is inherited from the channel estimation phase through the MRC receiver, is not as significant as the one described in Corollary~\ref{cor:decoding-lim_rho}. In fact, the normalized variance of the estimated symbols roughly decreases with $\rho$; on the other hand, the difference in amplitude between symbols cannot be recovered if $\rho$ is too high. Lastly, while for the channel estimation a reduction of the transmit SNR can be compensated by increasing the pilot length (see Section~\ref{sec:PA_EST_ub}), a low transmit SNR in the uplink data transmission phase inevitably results in a high normalized variance of the estimated symbols (see, e.g., Fig.~\ref{fig:var_VS_rho}).\footnote{For instance, assuming $s_{\ell} = 1$, it is straightforward to observe from \eqref{eq:s_var_norm-p=1} that $\lim_{\rho \to 0} \frac{\Vl}{|\El|^{2}} = \infty$.} In this respect, we point out that the system performance can be further enhanced by optimizing the transmit SNR separately for the two phases of channel estimation and uplink data transmission.

\begin{figure*}[t!]
\centering
\begin{subfigure}{0.325\textwidth}
\centering
\includegraphics[scale=1]{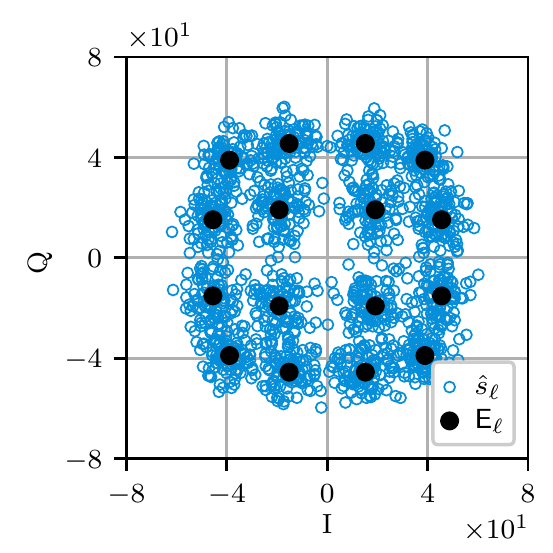}
\vspace{-0.5mm} \caption{$M = 64$, $\rho = 0$~dB, $\tau = 32$.}
\end{subfigure}
\begin{subfigure}{0.325\textwidth}
\centering
\includegraphics[scale=1]{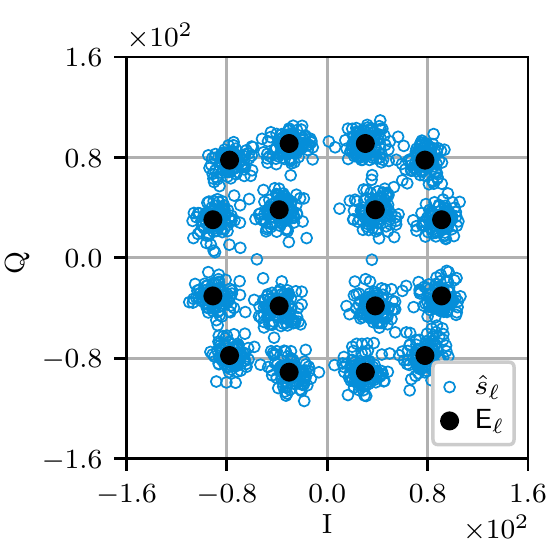}
\vspace{-0.5mm} \caption{$M = 128$, $\rho = 0$~dB, $\tau = 32$.}
\end{subfigure}
\begin{subfigure}{0.325\textwidth}
\centering
\includegraphics[scale=1]{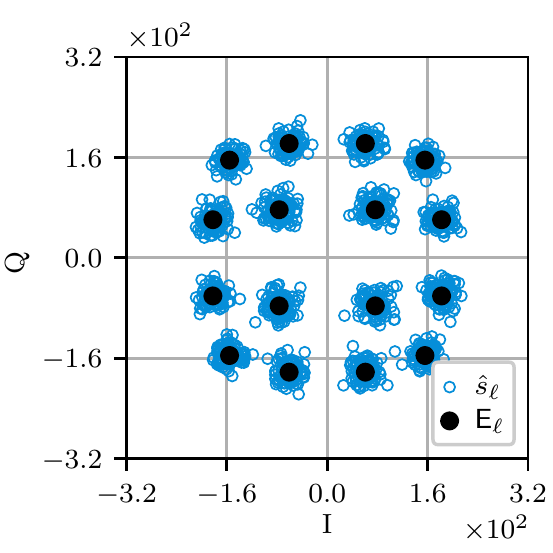}
\vspace{-0.5mm} \caption{$M = 256$, $\rho = 0$~dB, $\tau = 32$.}
\end{subfigure} \\
\begin{subfigure}{0.325\textwidth}
\centering
\includegraphics[scale=1]{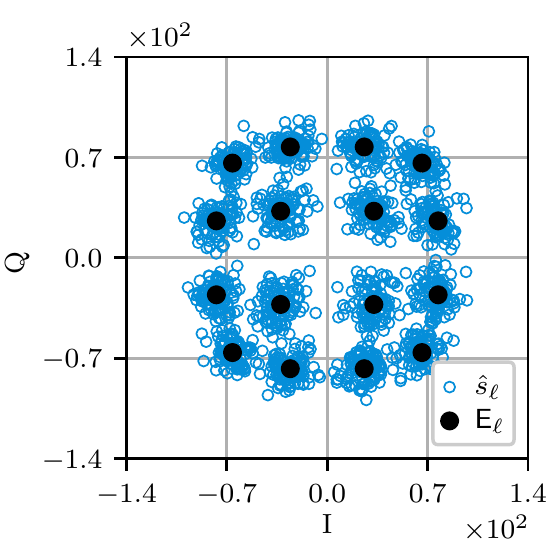}
\vspace{-0.5mm} \caption{$M = 128$, $\rho = 0$~dB, $\tau = 8$.}
\end{subfigure}
\begin{subfigure}{0.325\textwidth}
\centering
\includegraphics[scale=1]{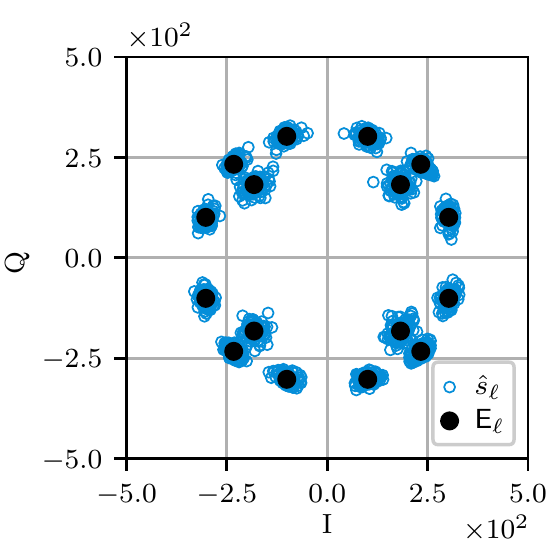}
\vspace{-0.5mm} \caption{$M = 128$, $\rho = 10$~dB, $\tau = 32$.}
\end{subfigure}
\begin{subfigure}{0.325\textwidth}
\centering
\includegraphics[scale=1]{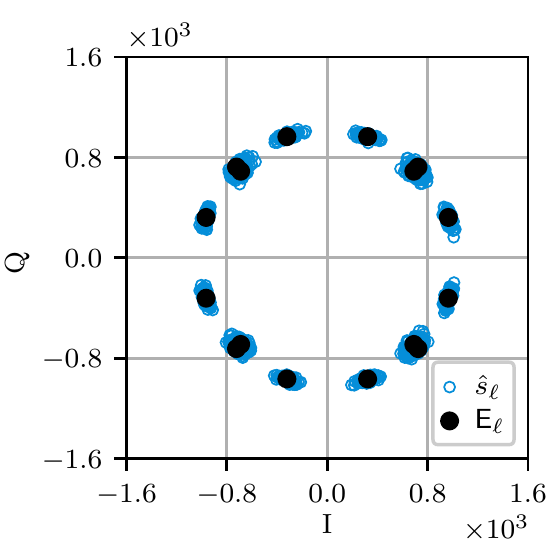}
\vspace{-0.5mm} \caption{$M = 128$, $\rho = 20$~dB, $\tau = 32$.}
\end{subfigure}
\caption{Estimated symbols with the MRC receiver, with 16-QAM transmit symbols and $K = 1$. The expected value of the estimated symbols is computed as in \eqref{eq:s_mean}.} \label{fig:dec_scatter} \vspace{-5.5mm}
\end{figure*}

\subsection{Numerical Results and Discussion} \label{sec:PA_DEC_num}

We now focus on the performance evaluation of the data detection with 1-bit ADCs with respect to the different parameters using the analytical results presented in Sections~\ref{sec:PA_DEC_main} and~\ref{sec:PA_DEC_ub}. In this regard, we assume that the BS uses the BLM estimator in \eqref{eq:H_hat_blm}, which is equivalent to the optimal scaled LS estimator in \eqref{eq:H_hat_sls_prime} when $K=1$, in the channel estimation phase and the MRC receiver in the data detection phase. We thus consider the expressions of $\El$ and $\Vl$ derived in \eqref{eq:s_mean} and \eqref{eq:s_var}, respectively, for the single-UE case (i.e, $K = 1$). As in Section~\ref{sec:PA_EST_num}, we use the pilots $\p = \p^{\star}$, with $\p^{\star}$ defined in \eqref{eq:p^star}, and $\p = \1_{\tau}$. Moreover, we specifically analyze the scenario where the set of transmit symbols $\setS$ corresponds to the 16-QAM constellation, i.e., $\setS = \frac{1}{\sqrt{10}} \big\{ \pm 1 \pm j, \pm 1 \pm j \, 3, \pm 3 \pm j, \pm 3 \pm j \, 3 \big\}$, which is normalized such that $\frac{1}{L} \sum_{\ell=1}^{L} |s_{\ell}|^{2} = 1$; however, we remark that our analytical framework is valid for any choice of $\setS$.

\begin{figure}[t!]
\centering
\includegraphics[scale=1]{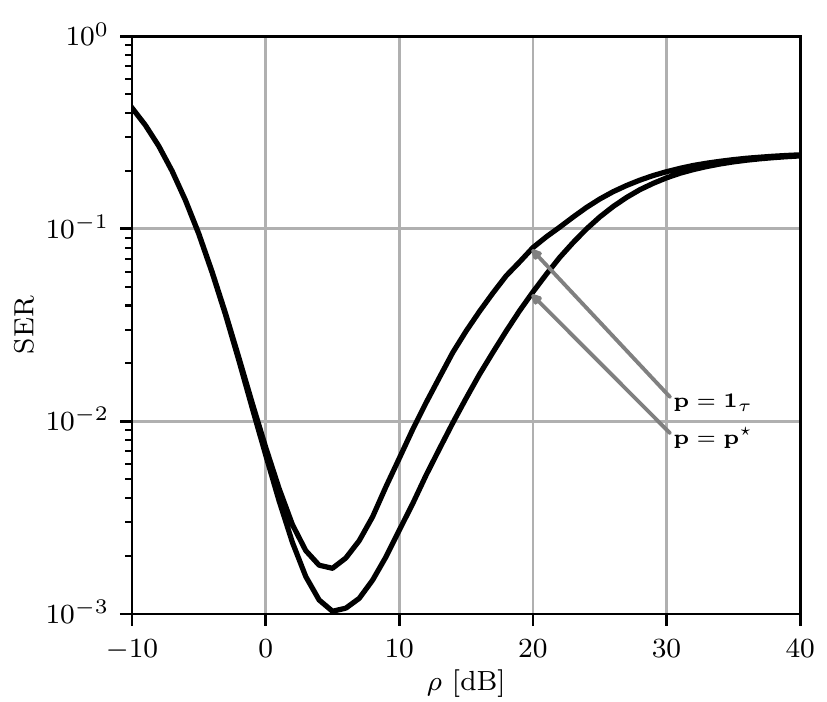}
\caption{SER against the transmit SNR, with 16-QAM transmit symbols, $K = 1$, $M = 128$, and $\tau = 32$. \\ } \label{fig:SER_VS_rho} \vspace{-5.5mm}
\end{figure}

Fig.~\ref{fig:dec_scatter} plots the estimated symbols for different settings, where each 16-QAM symbol is transmitted over $10^{2}$ independent channel realizations and $\p = \p^{\star}$ is used in the channel estimation phase. The expected value of the estimated symbols is computed as in Theorem~\ref{thm:s_mean}: this matches the sample average of the estimated symbols for each 16-QAM transmit symbol and can be used to efficiently implement MDD. Comparing Fig.~\ref{fig:dec_scatter}(a)--(c), which consider the same transmit SNR and pilot length, the relative dispersion of the estimated symbols about their expected value reduces as the number of BS antennas grows from $M = 64$ to $M = 256$. In fact, a higher granularity in the antenna domain allows to sum the contribution of a larger number of independent channel entries. On the other hand, comparing Fig.~\ref{fig:dec_scatter}(b) and~(d), which consider the same number of BS antennas and transmit SNR, the relative dispersion of the estimated symbols about their expected value slightly intensifies as we decrease the pilot length from $\tau = 32$ to $\tau = 8$. This stems from the overall diminished accuracy of the channel estimate used to compute the MRC receiver for each channel realization. Lastly, comparing Fig.~\ref{fig:dec_scatter}(b) and~(e)--(f), which consider the same number of BS antennas and pilot length, the estimated symbols resulting from the 16-QAM transmit symbols with the same phase, i.e., $\pm \frac{1}{\sqrt{10}} (1 \pm j)$ and $\pm \frac{1}{\sqrt{10}} (3 \pm j \, 3)$, get closer as the transmit SNR increases from $\rho = 0$~dB to $\rho = 10$~dB and they almost fully overlap when $\rho = 20$~dB. This behavior was observed in \cite{Jac17} and is formalized in Corollary~\ref{cor:decoding-lim_rho}, according to which such estimated symbols become identical in terms of both their expected value and variance at high SNR. In this respect, the SNR trade-off described in Sections~\ref{sec:PA_DEC_main} and~\ref{sec:PA_DEC_ub} is quite evident: while the normalized variance of the estimated symbols roughly decreases with $\rho$, the difference in amplitude between symbols cannot be recovered if $\rho$ is too high. For the 16-QAM, this produces a SER of about $25\%$ since there are four pairs of indistinguishable estimated symbols (see also Fig.~\ref{fig:SER_VS_rho}). In summary, having independent phases between the channel entries and operating at the right noise level are crucial to accurately estimate the phases and the amplitudes, respectively; we refer to Appendix~\ref{app:ch_est} and to the related discussion in \cite{Jac17} for more details.

\begin{figure}[t!]
\centering
\includegraphics[scale=1]{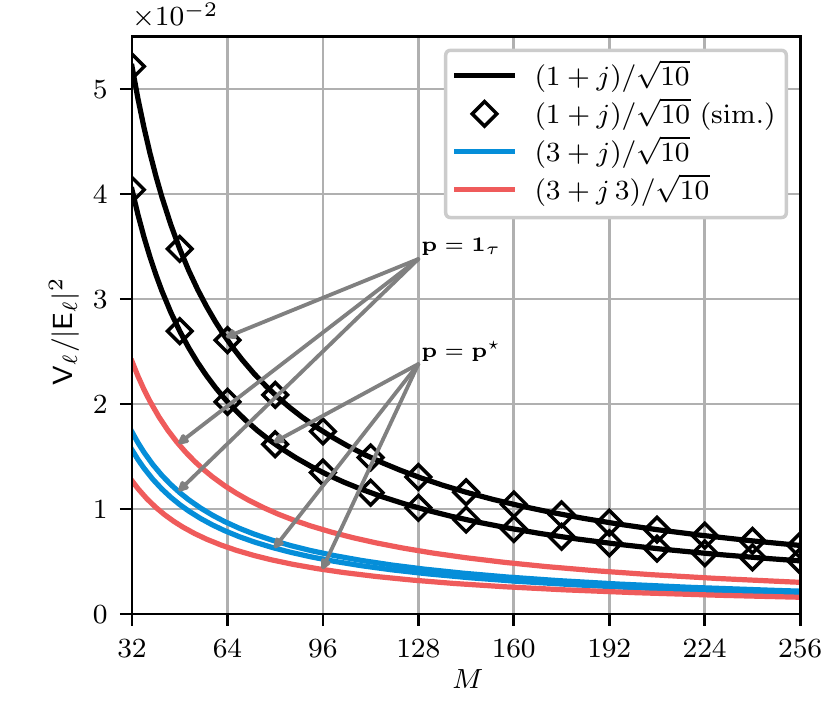}
\caption{Normalized variance of the estimated symbols against the number of BS antennas, with 16-QAM transmit symbols, $K = 1$, $\rho = 10$~dB, and $\tau = 32$.} \label{fig:var_VS_M} \vspace{-5.5mm}
\end{figure}

\begin{figure}[t!]
\centering
\includegraphics[scale=1]{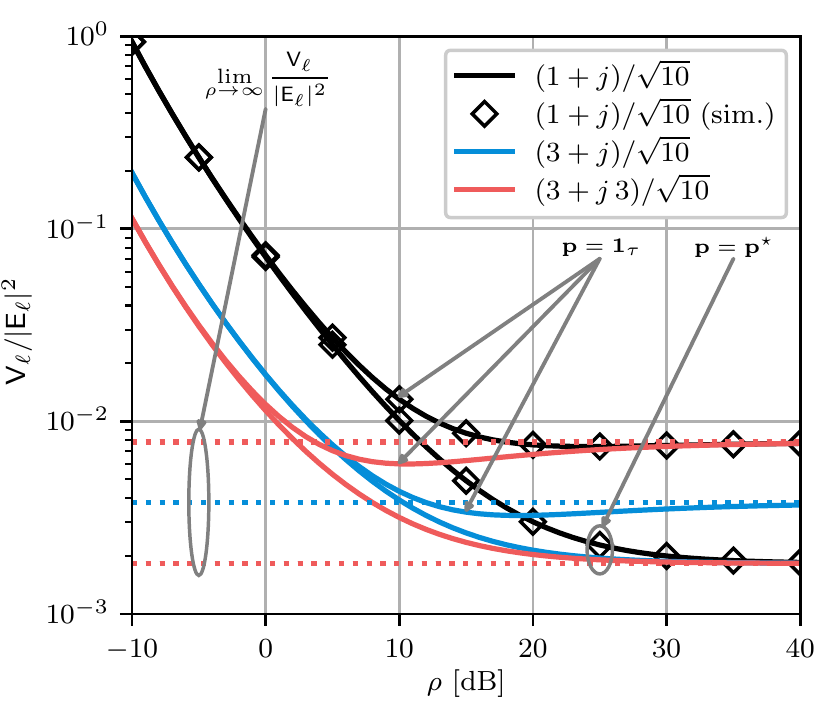}
\caption{Normalized variance of the estimated symbols against the transmit SNR, with 16-QAM transmit symbols, $K = 1$, $M = 128$, and $\tau = 32$.} \label{fig:var_VS_rho} \vspace{-4.5mm}
\end{figure}

\begin{figure}[t!]
\centering
\includegraphics[scale=1]{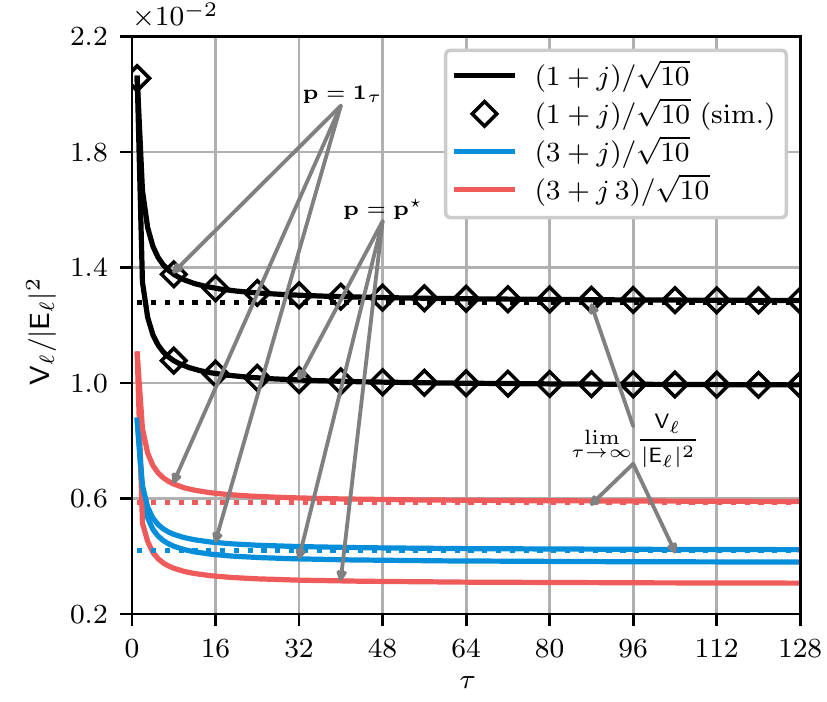}
\caption{Normalized variance of the estimated symbols against the pilot length, with 16-QAM transmit symbols, $K = 1$, $M = 128$, and $\rho = 10$~dB.} \label{fig:var_VS_tau} \vspace{-4mm}
\end{figure}

Let us now examine the behavior of the variance of the estimated symbols derived in Theorem~\ref{thm:s_var}, which we compare with Monte Carlo simulations with $10^{6}$ independent channel realizations. Fig.~\ref{fig:var_VS_M} considers $\rho = 10$~dB and $\tau = 32$, showing how the normalized variance of the estimated symbols $\frac{\Vl}{|\El|^{2}}$ decreases with the number of BS antennas $M$. The transmit symbols $\pm \frac{1}{\sqrt{10}} (1 \pm j)$, having the smallest power within the 16-QAM constellation, exhibit the most severe dispersion of the estimated symbols about their expected value. Furthermore, the upper bound in \eqref{eq:s_var_norm-p=1} becomes more accurate as $M$ grows. Fig.~\ref{fig:var_VS_rho} considers $M = 128$ and $\tau = 32$, showing that $\frac{\Vl}{|\El|^{2}}$ generally diminishes with the transmit SNR $\rho$ except for the SNR trade-off exhibited with $\p = \1_{\tau}$; here, the asymptotic expressions in \eqref{eq:s_var-lim_rho} and \eqref{eq:s_var_norm-lim_rho-p=1} are also included. Despite this trend, we recall that the difference in amplitude between symbols cannot be recovered if $\rho$ is too high, as discussed in the previous paragraph for Fig.~\ref{fig:dec_scatter}: thus, arbitrarily increasing the transmit SNR is detrimental for the system performance. Lastly, Fig.~\ref{fig:var_VS_tau} considers $M = 128$ and $\rho = 10$~dB, showing how $\frac{\Vl}{|\El|^{2}}$ reduces with the pilot length $\tau$.


We conclude this section by investigating the combined impact of the channel estimation and the data detection with 1-bit ADCs on the system performance in terms of SER, which we compute numerically via Monte Carlo simulations with $10^{6}$ independent channel realizations. In this context, the symbols are decoded by means of MDD aided by the result of Theorem~\ref{thm:s_mean}. Fig.~\ref{fig:SER_VS_rho} illustrates the SER against the transmit SNR $\rho$, with $M = 128$ and $\tau = 32$. Here, the SNR trade-off appears quite evident, whereby the SER decreases until it reaches its minimum at about $\rho = 5$~dB (where the upper bound obtained with $\p = \1_{\tau}$ proves to be remarkably tight) before escalating again. Then, the SER asymptotically reaches $25\%$ at high SNR, where the inner estimated symbols of the 16-QAM constellation become indistinguishable from the outer estimated symbols with the same phase (see also Fig.~\ref{fig:dec_scatter}(f)). We remark that the SER can be further reduced by optimizing the transmit SNR separately for the two phases of channel estimation and uplink data transmission, which is left~for~future~work.


\begin{figure*}[t!]
\addtocounter{equation}{+2}
\begin{align}
\label{eq:ch_est2} Q(h_{m} p_{u}^{*}) p_{u} & = Q \big( e^{j \, (\theta_{m} - \phi_{u})} \big) e^{j \, \phi_{u}} \\
\label{eq:ch_est3} & = \left\{
\begin{array}{ll}
\sqrt{\frac{\rho + 1}{2}} \big( \sgn \big( \Re[h_{m}] \big) + j \, \sgn \big( \Im[h_{m}] \big) \big) e^{j \, \phi_{u}}	& \textrm{if}~\phi_{u} \in \big[ \vartheta_{m} - \frac{\pi}{2}, \vartheta_{m} \big], \\
\sqrt{\frac{\rho + 1}{2}} \big( \sgn \big( \Im[h_{m}] \big) - j \, \sgn \big( \Re[h_{m}] \big) \big) e^{j \, \phi_{u}}	& \textrm{if}~\phi_{u} \in \big[ \vartheta_{m}, \vartheta_{m} + \frac{\pi}{2} \big], \\
\sqrt{\frac{\rho + 1}{2}} \big( - \sgn \big( \Re[h_{m}] \big) - j \, \sgn \big( \Im[h_{m}] \big) \big) e^{j \, \phi_{u}}	& \textrm{if}~\phi_{u} \in \big[ \vartheta_{m} + \frac{\pi}{2}, \vartheta_{m} + \pi \big], \\
\sqrt{\frac{\rho + 1}{2}} \big( - \sgn \big( \Im[h_{m}] \big) + j \, \sgn \big( \Re[h_{m}] \big) \big) e^{j \, \phi_{u}}	& \textrm{if}~\phi_{u} \in \big[ \vartheta_{m} + \pi, \vartheta_{m} + \frac{3 \pi}{2} \big]
\end{array} \right.
\end{align}
\hrulefill
\begin{align}
\nonumber \lim_{\tau \to \infty} \frac{1}{\tau} \sum_{u=1}^{\tau} Q(h_{m} p_{u}^{*}) p_{u} & = \frac{1}{2 \pi} \sqrt{\frac{\rho + 1}{2}} \bigg( \Big( \sgn \big( \Re[h_{m}] \big) + j \, \sgn \big( \Im[h_{m}] \big) \Big) \int_{\vartheta_{m} - \frac{\pi}{2}}^{\vartheta_{m}} e^{j \, \phi} \diff \phi \\
\nonumber & \phantom{=} \ + \Big( \sgn \big( \Im[h_{m}] \big) - j \, \sgn \big( \Re[h_{m}] \big) \Big) \int_{\vartheta_{m}}^{\vartheta_{m} + \frac{\pi}{2}} e^{j \, \phi} \diff \phi + \Big( - \sgn \big( \Re[h_{m}] \big) - j \, \sgn \big( \Im[h_{m}] \big) \Big) \\
\label{eq:ch_est4} & \phantom{=} \ \times \int_{\vartheta_{m} + \frac{\pi}{2}}^{\vartheta_{m} + \pi} e^{j \, \phi} \diff \phi + \Big( - \sgn \big( \Im[h_{m}] \big) + j \, \sgn \big( \Re[h_{m}] \big) \Big) \int_{\vartheta_{m} + \pi}^{\vartheta_{m} + \frac{3 \pi}{2}} e^{j \, \phi} \diff \phi \bigg) \\
\label{eq:ch_est5} & = \frac{2}{\pi} \sqrt{\frac{\rho + 1}{2}} (1 - j) \Big( \sgn \big( \Re[h_{m}] \big) + j \, \sgn \big( \Im[h_{m}] \big) \Big) e^{j \, \vartheta_{m}}
\end{align}
\hrulefill \vspace{-4mm}
\end{figure*}

\section{Conclusions} \label{sec:CONCL}

This paper presents an analytical framework for the channel estimation and the data detection in massive MIMO uplink systems with 1-bit ADCs. First, we provide a precise characterization of the MSE of the channel estimation with respect to different parameters. In addition, we show that, for i.i.d. Rayleigh fading, the BLM estimator can be simplified as a scaled LS estimator with UE-specific scaling factors and that using a common optimized scaling factor for all the UEs entails no noticeable performance loss. For the data detection, we characterize the expected value and the variance of the estimated symbols when MRC is adopted. These results can be exploited to efficiently implement MDD and to properly design the set of transmit symbols. The proposed analysis gives important practical insights into the design and the implementation of 1-bit quantized systems. In particular, it highlights a fundamental SNR trade-off, according to which arbitrarily increasing the transmit SNR is detrimental for the system performance. In this respect, the optimal transmit SNR for the channel estimation is shown to decrease as the pilot length increases.

Future work will consider extensions of the proposed analytical framework to more realistic channel models (for the channel estimation) and to the multi-UE case (for the data detection), as well as a SER optimal design of the set of transmit symbols capitalizing on our data detection analysis.

\section*{Acknowledgments}

The authors would like to thank the anonymous reviewers, whose comments and suggestions helped to improve the paper.

\appendices

\section{Fundamentals of Channel Estimation with 1-bit ADCs} \label{app:ch_est}

Assuming $K=1$, let $\h \triangleq (h_{m}) \in \Compl^{M \times 1}$ and $\p \triangleq (p_{u}) \in \Compl^{\tau \times 1}$ denote the uplink channel vector and the pilot, respectively, of the UE. When a scaled LS estimator (such as the one in \eqref{eq:H_hat_sls}) is used, the channel estimate $\hat{\h} \triangleq (\hat{h}_{m})$ is obtained as
\addtocounter{equation}{-6}
\begin{align}
\label{eq:ch_est1} \hat{\h} & = \sqrt{\Psi} Q \left( \! \begin{bmatrix}
\sqrt{\rho} h_{1} p_{1}^{*} \! + \! Z_{1,1}	& \hspace{-3mm} \cdots	& \hspace{-3mm} \sqrt{\rho} h_{1} p_{\tau}^{*} \! + \! Z_{1,\tau} \\
\vdots								& \hspace{-3mm} \ddots 	& \hspace{-3mm} \vdots \\
\sqrt{\rho} h_{M} p_{1}^{*} \! + \! Z_{M,1}	& \hspace{-3mm} \cdots	& \hspace{-3mm} \sqrt{\rho} h_{M} p_{\tau}^{*} \! + \! Z_{M,\tau}
\end{bmatrix} \! \right) \! \begin{bmatrix}
p_{1} \\
\vdots \\
p_{\tau}
\end{bmatrix}
\end{align}
with
\begin{align}
\nonumber \hat{h}_{m} & = \sqrt{\frac{\rho + 1}{2} \Psi} \sum_{u=1}^{\tau} p_{u} \Big( \sgn \big( \Re[\sqrt{\rho} h_{m} p_{u}^{*} + Z_{m,u}] \big) \\
& \phantom{=} \ + j \, \sgn \big( \Im[\sqrt{\rho} h_{m} p_{u}^{*} + Z_{m,u}] \big) \Big).
\end{align}

Let $h_{m} = \alpha_{m} e^{j \, \theta_{m}}$, with $\vartheta_{m} \triangleq \big( \theta_{m}~\mathrm{mod}~\frac{\pi}{2} \big)$, and let $p_{u} = e^{j \, \phi_{u}}$ (recall that $|p_{u}|^{2} = 1$, $\forall u$). Assuming $\rho \to \infty$, the phase of $h_{m}$ can be estimated from $Q(h_{m} \p^{\herm}) \p$ as detailed in \eqref{eq:ch_est2}--\eqref{eq:ch_est3} at the top of the page, i.e., $Q \big( e^{j \, (\theta_{m} - \phi_{u})} \big)$ shifts quadrant according to the phase of $p_{u}$. Assuming that the entries of $\p$ span the unit circle, in the limit of $\tau \to \infty$, we obtain \eqref{eq:ch_est4}--\eqref{eq:ch_est5} at the top of the page, where \eqref{eq:ch_est5} follows from
\addtocounter{equation}{+4}
\begin{align}
\int_{\vartheta_{m} - \frac{\pi}{2}}^{\vartheta_{m}} e^{j \, \phi} \diff \phi & = (1 - j) e^{j \, \vartheta_{m}}, \\
\int_{\vartheta_{m}}^{\vartheta_{m} + \frac{\pi}{2}} e^{j \, \phi} \diff \phi & = (1 + j) e^{j \, \vartheta_{m}}, \\
\int_{\vartheta_{m} + \frac{\pi}{2}}^{\vartheta_{m} + \pi} e^{j \, \phi} \diff \phi & = (-1 + j) e^{j \, \vartheta_{m}}, \\
\int_{\vartheta_{m} + \pi}^{\vartheta_{m} + \frac{3 \pi}{2}} e^{j \, \phi} \diff \phi & = (-1 - j) e^{j \, \vartheta_{m}}.
\end{align}
Finally, from \eqref{eq:ch_est5}, we have
\begin{align}
\nonumber & (1 - j) \Big( \sgn \big( \Re[h_{m}] \big) + j \, \sgn \big( \Im[h_{m}] \big) \Big) e^{j \, \vartheta_{m}} \\
& = \left\{
\begin{array}{ll}
2 e^{j \, \vartheta_{m}}			& \textrm{if}~\theta_{m} \in \big[ 0, \frac{\pi}{2} \big]~(\textrm{i.e.,~if}~\theta_{m} = \vartheta_{m}), \\
2 j \, e^{j \, \vartheta_{m}}		& \textrm{if}~\theta_{m} \in \big[ \frac{\pi}{2}, \pi \big]~(\textrm{i.e.,~if}~\theta_{m} = \vartheta_{m} + \frac{\pi}{2}), \\
- 2 e^{j \, \vartheta_{m}}		& \textrm{if}~\theta_{m} \in \big[ \pi, \frac{3 \pi}{2} \big]~(\textrm{i.e.,~if}~\theta_{m} = \vartheta_{m} + \pi), \\
- 2 j \, e^{j \, \vartheta_{m}}	& \textrm{if}~\theta_{m} \in \big[ \frac{3 \pi}{2}, 2 \pi \big]~(\textrm{i.e.,~if}~\theta_{m} = \vartheta_{m} + \frac{3 \pi}{2})
\end{array} \right. \\
& = 2 e^{j \, \theta_{m}}
\end{align}
which yields
\begin{align} \label{eq:ch_est6}
\lim_{\tau \to \infty} \frac{1}{\tau} \sum_{u=1}^{\tau} Q(h_{m} p_{u}^{*}) p_{u} & = \frac{4}{\pi} \sqrt{\frac{\rho + 1}{2}} e^{j \, \theta_{m}}.
\end{align}
Hence, the phase of $h_{m}$ can be estimated accurately if $\tau$ is sufficiently large and the pilot symbols span the unit circle. Nonetheless, from \eqref{eq:ch_est2}--\eqref{eq:ch_est3}, it is straightforward to see that $Q \big( e^{j \, (\theta_{m} - \phi_{u})} \big) e^{j \, \phi_{u}} = Q \big( e^{j \, (\theta_{m} - \phi_{u} \mp \frac{\pi}{2})} \big) e^{j \, (\phi_{u} \pm \frac{\pi}{2})} = Q \big( e^{j \, (\theta_{m} - \phi_{u} \mp \pi)} \big) e^{j \, (\phi_{u} \pm \pi)}$, i.e., shifting the phase of the pilot symbol by a multiple of $\frac{\pi}{2}$ does not add any information about the phase of $h_{m}$ when $\rho \to \infty$. As a consequence, the best possible pilot choice features equispaced and non-repeating phases on an interval $\big[ \eta, \eta + \frac{\pi}{2} \big]$, with $\eta \in [0, 2 \pi]$ (one such choice is $\p^{\star}$ in \eqref{eq:p^star}). On the other hand, the worst possible pilot choice is given by fixing $\p$ such that $p_{u} \in \{\pm \beta, \pm j \, \beta\}$, $\forall u$, with $\beta \in \Compl$ and $|\beta|^{2} = 1$ (one such choice is $\p = \1_{\tau}$). Note that fixing $\p = \1_{\tau}$ with $\rho \to \infty$ would reduce each channel entry to a scaled symbol of the QPSK constellation regardless of the value of $\tau$, whereas with finite $\rho$ the phase of $h_{m}$ can be still estimated by exploiting the independent noise realizations over the pilot symbols.

The right-hand side of \eqref{eq:ch_est6} does not include any information about the amplitude of $h_{m}$ due to the assumption that $\rho \to \infty$. Assuming now finite $\rho$ and, for simplicity, $\p = \1_{\tau}$, the amplitude of $h_{m}$ can be estimated from $Q \big( \sqrt{\rho} h_{m} \1_{\tau}^{\tran} + [Z_{m,1}, \ldots, Z_{m,\tau}] \big) \1_{\tau}$, where
\begin{align}
\nonumber Q(\sqrt{\rho} h_{m} + Z_{m,u}) & = \sqrt{\frac{\rho + 1}{2}} \Big( \sgn \big( \sqrt{\rho} \Re[h_{m}] + \Re[Z_{m,u}] \big) \\
& \phantom{=} \ + j \, \sgn \big( \sqrt{\rho} \Im[h_{m}] + \Im[Z_{m,u}] \big) \Big).
\end{align}
In the limit of $\tau \to \infty$, we have
\begin{align}
\nonumber \hspace{-1mm} \lim_{\tau \to \infty} \frac{1}{\tau} \sum_{u=1}^{\tau} Q(\sqrt{\rho} h_{m} + Z_{m,u}) & = \sqrt{\frac{\rho + 1}{2}} \Big( \erf \big( \sqrt{\rho} \Re[h_{m}] \big) \\
& \phantom{=} \ + j \, \erf \big( \sqrt{\rho} \Im[h_{m}] \big) \Big)
\end{align}
where $\erf(w) \triangleq \frac{2}{\sqrt{\pi}} \int_{0}^{w} e^{- t^{2}} \diff t$ denotes the error function. Since $\erf(w)$ is approximately linear for $w \in [-1, 1]$, the difference in amplitude between channel entries can be estimated accurately if their real and imaginary parts lie in $\big[ - \frac{1}{\sqrt{\rho}}, \frac{1}{\sqrt{\rho}} \big]$ and $\tau$ is sufficiently large, and this holds despite choosing $\p = \1_{\tau}$. On the other hand, if $\tau$ is not sufficiently large at low SNR, the channel estimates are corrupted by the strong noise. Hence, the estimation of the amplitude of $h_{m}$ benefits from operating at the right noise level.

\section{Proof of Theorem~\ref{thm:MSE_blm}} \label{app:MSE_blm}

We begin by writing \eqref{eq:MSE_blm_0} as
\begin{align}
\mse_{\blm} & = \frac{1}{M K} \Big( \! \Exp [\underline{\h}^{\herm} \underline{\h}] \! + \! \Exp [\underline{\hat{\h}}_{\blm}^{\herm} \underline{\hat{\h}}_{\blm}] \! - \! 2 \Exp \big[ \Re [\underline{\hat{\h}}_{\blm}^{\herm} \underline{\h}] \big] \! \Big) \\
\nonumber & = 1 + \frac{1}{M K} \bigg( \frac{2}{\pi} \rho \tr \big( \tilde{\P}^{\tran} \Sigmab_{\rmp}^{-1} \Exp[\r_{\rmp} \r_{\rmp}^{\herm}] \Sigmab_{\rmp}^{-1} \tilde{\P}^{*} \big) \\
& \phantom{=} \ - 2 \sqrt{\frac{2}{\pi} \rho} \Re \Big[ \tr \big( \Exp[\underline{\h} \r_{\rmp}^{\herm}] \Sigmab_{\rmp}^{-1} \tilde{\P}^{*} \big) \Big] \bigg) \\
\label{eq:MSEp_blm1} & = 1 - \frac{1}{M K} \frac{2}{\pi} \rho \tr(\tilde{\P}^{\tran} \Sigmab_{\rmp}^{-1} \tilde{\P}^{*})
\end{align}
where \eqref{eq:MSEp_blm1} follows from applying $\Exp[\underline{\h} \r_{\rmp}^{\herm}] = \sqrt{\frac{2}{\pi} \rho} \tilde{\P}^{\tran}$; note that a similar MSE expression appears in \cite[Eq.~(48)]{Rao21}. Now, as detailed in Appendix~\ref{app:Sigmap}, we can express the covariance matrix of $\underline{\r}_{\rmp}$ as
\begin{align} \label{eq:Sigmap}
\Sigmab_{\rmp} & = (\rho K + 1) \Phib \otimes \I_{M}
\end{align}
where we have defined $\Phib \triangleq (\Phi_{u,v}) \in \Compl^{\tau \times \tau}$, with
\begin{align}
\Phi_{u,v} & \triangleq \left\{
\begin{array}{ll}
1 & \hspace{-17mm} \textrm{if}~u = v, \\
\Omega \Big( \frac{\rho \sum_{k=1}^{K} \Re[P_{u,k} P_{v,k}^{*}]}{\rho K + 1} \Big) \! - \! j \, \Omega \Big( \frac{\rho \sum_{k=1}^{K} \Im[P_{u,k} P_{v,k}^{*}]}{\rho K + 1} \Big) & \\
& \hspace{-17mm} \textrm{if}~u \neq v.
\end{array} \right.
\end{align}
Furthermore, let $\p_{k} \in \Compl^{\tau \times 1}$ denote the $k$th column of $\P$. Hence, we can write

\newpage

$ $ \vspace{-8mm}

\begin{align}
\tilde{\P}^{\tran} \Sigmab_{\rmp}^{-1} \tilde{\P}^{*} & = \frac{1}{\rho K + 1} (\P^{\tran} \Phib^{-1} \P^{*}) \otimes \I_{M} \\
\label{eq:MSEp_blm2} & = \frac{\tau^{2}}{\rho K + 1} (\P^{\tran} \Phib \P^{*})^{-1} \otimes \I_{M}
\end{align}
where \eqref{eq:MSEp_blm2} results from the fact that, if $\P$ is chosen such that $\P \P^{\herm}$ is circulant, $\Phib$ is also circulant and, as a consequence, so is its inverse: in this case, $\P$ diagonalizes both $\Phib$ and its inverse, which implies $\p_{k}^{\tran} \Phib \p_{i}^{*} = 0$, $\forall k \neq i$. Finally, plugging \eqref{eq:MSEp_blm2} into \eqref{eq:MSEp_blm1} yields
\begin{align}
\hspace{-1mm} \mse_{\blm} & = 1 - \frac{1}{M K} \frac{2}{\pi} \frac{\rho \tau^{2}}{\rho K + 1} \tr \big( (\P^{\tran} \Phib \P^{*})^{-1} \otimes \I_{M} \big) \\
& = 1 - \frac{1}{K} \frac{2}{\pi} \frac{\rho \tau^{2}}{\rho K + 1} \tr \big( (\P^{\tran} \Phib \P^{*})^{-1} \big) \\
& = 1 - \frac{2}{\pi} \frac{\rho \tau^{2}}{\rho K + 1} \frac{1}{K} \sum_{k=1}^{K} \frac{1}{\p_{k}^{\tran} \Phib \p_{k}^{*}}
\end{align}
and the expression in \eqref{eq:MSE_blm} is obtained by observing that $\p_{k}^{\tran} \Phib \p_{k}^{*} = \tau + \delta_{k}$, with $\delta_{k}$ defined in \eqref{eq:delta_k}. \hfill \IEEEQED

\subsection{Derivations of \eqref{eq:Sigmap}} \label{app:Sigmap}

In this section, we derive the closed-form expression of $\Sigmab_{\rmp}$. To this end, we introduce the following definitions:
\begin{align}
\label{eq:A1} A_{m,u} & \triangleq \sgn \bigg( \Re \bigg[ \sqrt{\rho} \sum_{k=1}^{K} H_{m,k} P_{u,k}^{*} + Z_{m,u} \bigg] \bigg) \\
\nonumber & = \sgn \bigg( \sqrt{\rho} \sum_{k=1}^{K} \big( \Re[H_{m,k}] \Re[P_{u,k}] \\
\label{eq:A2} & \phantom{=} \ + \Im[H_{m,k}] \Im[P_{u,k}] \big) + \Re[Z_{m,u}] \bigg), \\
\label{eq:B1} B_{m,u} & \triangleq \sgn \bigg( \Im \bigg[ \sqrt{\rho} \sum_{k=1}^{K} H_{m,k} P_{u,k}^{*} + Z_{m,u} \bigg] \bigg) \\
\nonumber & = \sgn \bigg( \sqrt{\rho} \sum_{k=1}^{K} \big( - \Re[H_{m,k}] \Im[P_{u,k}] \\
\label{eq:B2} & \phantom{=} \ + \Im[H_{m,k}] \Re[P_{u,k}] \big) + \Im[Z_{m,u}] \bigg).
\end{align}
Moreover, we present the following proposition, which will be also used in Appendix~\ref{app:s_mean}.

\begin{figure*}
\addtocounter{equation}{+5}
\begin{align} \label{eq:RR_exp}
\Exp [R_{m,u} R_{n,v}^{*}] & = \left\{
\begin{array}{ll}
(\rho K + 1) \Big( \Omega \Big( \frac{\rho \sum_{k=1}^{K} \Re[P_{u,k} P_{v,k}^{*}]}{\rho K + 1} \Big) - j \, \Omega \Big( \frac{\rho \sum_{k=1}^{K} \Im[P_{u,k} P_{v,k}^{*}]}{\rho K + 1} \Big) \Big) & \textrm{if}~m = n, u \neq v, \\
0 & \textrm{if}~m \neq n
\end{array} \right.
\end{align}
\hrulefill
\addtocounter{equation}{+3}
\begin{align}
\label{eq:big_exp1} \zetab & = \big[ \Re[H_{m,1}], \ldots, \Re[H_{m,K}], \Im[H_{m,1}], \ldots, \Im[H_{m,K}], \Re[Z_{m,u}], \Re[Z_{m,v}] \big]^{\tran} \sim \setN \bigg( \0_{(2K+2)}, \frac{1}{2} \I_{(2K+2)} \bigg), \\
\a_{1} & = \big[ \sqrt{\rho} \Re[P_{u,1}], \ldots, \sqrt{\rho} \Re[P_{u,K}], \sqrt{\rho} \Im[P_{u,1}], \ldots, \sqrt{\rho} \Im[P_{u,K}], 1, 0 \big]^{\tran} \in \Real^{(2K+2) \times 1}, \\
\label{eq:big_exp2} \a_{2} & = \big[ \sqrt{\rho} \Re[P_{v,1}], \ldots, \sqrt{\rho} \Re[P_{v,K}], \sqrt{\rho} \Im[P_{v,1}], \ldots, \sqrt{\rho} \Im[P_{v,K}], 0, 1 \big]^{\tran} \in \Real^{(2K+2) \times 1}
\end{align}

\vspace{-2.5mm}

\hrulefill
\begin{align}
\nonumber & \Exp \bigg[ \sgn \bigg( \sqrt{\rho} \sum_{i=1}^{K} \big( \Re[H_{m,i}] \Re[P_{u,i}] + \Im[H_{m,i}] \Im[P_{u,i}] \big) + \Re[Z_{m,u}] \bigg)  \\
\nonumber & \times \sgn \bigg( \sqrt{\rho} \sum_{i=1}^{K} \big( \Re[H_{m,i}] \Re[P_{v,i}] + \Im[H_{m,i}] \Im[P_{v,i}] \big) + \Re[Z_{m,v}] \bigg) \bigg] \\
\label{eq:big_exp3} & = \Omega \bigg( \frac{\rho \sum_{k=1}^{K} (\Re[P_{u,k}] \Re[P_{v,k}] + \Im[P_{u,k}] \Im[P_{v,k}])}{(\rho \sum_{k=1}^{K} (\Re[P_{u,k}]^{2} + \Im[P_{u,k}]^{2}) + 1)^{\frac{1}{2}} (\rho \sum_{k=1}^{K} (\Re[P_{v,k}]^{2} + \Im[P_{v,k}]^{2}) + 1)^{\frac{1}{2}}} \bigg)
\end{align}
\hrulefill \vspace{-4mm}
\end{figure*}

\begin{proposition} \label{pro:exp_sign} \rm{
Let $\zetab \sim \setN (0, \gamma \I_{N})$, with $\gamma > 0$. For $\a_{1}, \a_{2} \in \Real^{N \times 1}$, we have
\addtocounter{equation}{-13}
\begin{align} \label{eq:exp_sign}
\Exp \big[ \sgn(\a_{1}^{\tran} \zetab) \sgn(\a_{2}^{\tran} \zetab) \big] = \Omega \bigg( \frac{\a_{1}^{\tran} \a_{2}}{\|\a_{1}\| \, \|\a_{2}\|} \bigg).
\end{align}}
\end{proposition}

\begin{IEEEproof}
To obtain the expression in \eqref{eq:exp_sign}, we first observe that
\begin{align}
\nonumber & \Exp \big[ \sgn (X_{1}) \sgn (X_{2}) \big] \\
\nonumber & = \Pr [X_{1} > 0, X_{2} > 0] + \Pr [X_{1} < 0, X_{2} < 0] \\
\label{eq:exp_sign1} & \phantom{=} \ - \Pr [X_{1} > 0, X_{2} < 0] - \Pr [X_{1} < 0, X_{2} > 0].
\end{align}
For $X_{1} = \a_{1}^{\tran} \zetab$ and $X_{2} = \a_{2}^{\tran} \zetab$, the first term on the right-hand side of \eqref{eq:exp_sign1} can be obtained building on \cite{Abr64} as
\begin{align}
\Pr \big[ \a_{1}^{\tran} \zetab > 0, \a_{2}^{\tran} \zetab > 0 \big] & = \frac{1}{4} \bigg( 1 + \Omega \bigg( \frac{\a_{1}^{\tran} \a_{2}}{\|\a_{1}\| \, \|\a_{2}\|} \bigg) \bigg)
\end{align}
where $\frac{\a_{1}^{\tran} \a_{2}}{\|\a_{1}\| \, \|\a_{2}\|}$ represents the correlation coefficient between $X_{1}$ and $X_{2}$, and the other terms can be derived following similar steps.
\end{IEEEproof} \vspace{1mm}

Now, we can write the $\big( (u-1)M+m,(v-1)M+n \big)$th entry of $\underline{\r}_{\rmp} \underline{\r}_{\rmp}^{\herm}$ as
\begin{align}
\hspace{-2mm} R_{m,u} R_{n,v}^{*} & = \frac{\rho K + 1}{2} (A_{m,u} + j \, B_{m,u})(A_{n,v} + j \, B_{n,v})^{*} \\
\nonumber & = \frac{\rho K + 1}{2} \big( A_{m,u} A_{n,v} + B_{m,u} B_{n,v} \\
\label{eq:RR} & \phantom{=} \ + j \, (B_{m,u} A_{n,v} - A_{m,u} B_{n,v}) \big)
\end{align}
with $R_{m,u} R_{m,u}^{*} = \rho K + 1$. Hence, the expected value of \eqref{eq:RR} is given by \eqref{eq:RR_exp} at the top of the page. This results from
\addtocounter{equation}{+1}
\begin{align}
\label{eq:A_A_exp1} \Exp [A_{m,u} A_{m,v}] & = \Exp [B_{m,u} B_{m,v}] \\
\label{eq:A_A_exp2} & = \Omega \bigg( \frac{\rho \sum_{k=1}^{K} \Re[P_{u,k} P_{v,k}^{*}]}{\rho K + 1} \bigg), \\
\label{eq:A_B_exp} \Exp [A_{m,u} B_{m,v}] & = \Omega \bigg( \frac{\rho \sum_{k=1}^{K} \Im[P_{u,k} P_{v,k}^{*}]}{\rho K + 1} \bigg)
\end{align}
which are derived by applying Proposition~\ref{pro:exp_sign}. For instance, $\Exp [A_{m,u} A_{m,v}]$ in \eqref{eq:A_A_exp1}--\eqref{eq:A_A_exp2} can be obtained by plugging \eqref{eq:big_exp1}--\eqref{eq:big_exp2} at the top of the page into \eqref{eq:exp_sign}, which gives \eqref{eq:big_exp3} at the top of the page. Finally, the expression in \eqref{eq:Sigmap} readily follows from \eqref{eq:RR_exp}.

\section{Proof of Corollary~\ref{cor:MSE_blm_sls}} \label{app:MSE_blm_sls}

We show that the estimator in \eqref{eq:H_hat_blm_sls} provides the same MSE of the channel estimation as the BLM estimator in \eqref{eq:H_hat_blm}. To this end, we write $\mathrm{vec}[\hat{\H}_{\blm}] = \tilde{\Psib}^{\frac{1}{2}} \tilde{\P}^{\tran} \underline{\r}_{\rmp}$, where we have defined $\tilde{\Psib} \triangleq \Psib \otimes \I_{M} \in \Compl^{M K \times M K}$. Following similar steps as in the proof of Theorem~\ref{thm:MSE_blm}, we obtain
\addtocounter{equation}{+4}
\begin{align}
\nonumber \mse_{\blm} & = 1 + \frac{1}{M K} \Big( \tr \big( \tilde{\Psib}^{\frac{1}{2}} \tilde{\P}^{\tran} \Exp[\r_{\rmp} \r_{\rmp}^{\herm}] \tilde{\P}^{*} \tilde{\Psib}^{\frac{1}{2}} \big) \\
& \phantom{=} \ - 2 \Re \Big[ \tr \big( \Exp[\underline{\h} \r_{\rmp}^{\herm}] \tilde{\P}^{*} \tilde{\Psib}^{\frac{1}{2}} \big) \Big] \Big) \\
\nonumber & = 1 + \frac{1}{M K} \bigg( \tr (\tilde{\Psib}^{\frac{1}{2}} \tilde{\P}^{\tran} \Sigmab_{\rmp} \tilde{\P}^{*} \tilde{\Psib}^{\frac{1}{2}}) \\
\label{eq:MSEp_blm_sls1} & \phantom{=} \ - 2 \sqrt{\frac{2}{\pi} \rho} \tr (\tilde{\P}^{\tran} \tilde{\P}^{*} \tilde{\Psib}^{\frac{1}{2}}) \bigg).
\end{align}
Hence, we can write

\newpage

$ $ \vspace{-8.5mm}

\begin{align}
\tilde{\Psib}^{\frac{1}{2}} \tilde{\P}^{\tran} \Sigmab_{\rmp} \tilde{\P}^{*} \tilde{\Psib}^{\frac{1}{2}} & = (\rho K + 1) (\Psib^{\frac{1}{2}} \P^{\tran} \Phib \P^{*} \Psib^{\frac{1}{2}}) \otimes \I_{M}, \\
\tilde{\P}^{\tran} \tilde{\P}^{*} \tilde{\Psib}^{\frac{1}{2}} & = (\P^{\tran} \P^{*} \Psib^{\frac{1}{2}}) \otimes \I_{M}
\end{align}
and \eqref{eq:MSEp_blm_sls1} becomes
\begin{align}
\nonumber \mse_{\blm} & = 1 + \frac{1}{M K} \bigg( (\rho K + 1) \tr \big( (\Psib^{\frac{1}{2}} \P^{\tran} \Phib \P^{*} \Psib^{\frac{1}{2}}) \otimes \I_{M} \big) \\
& \phantom{=} \ - 2 \sqrt{\frac{2}{\pi} \rho} \tr \big( (\P^{\tran} \P^{*} \Psib^{\frac{1}{2}}) \otimes \I_{M} \big) \bigg) \\
\nonumber & = 1 + \frac{1}{K} \bigg( (\rho K + 1) \tr (\Psib^{\frac{1}{2}} \P^{\tran} \Phib \P^{*} \Psib^{\frac{1}{2}}) \\
& \phantom{=} \ - 2 \sqrt{\frac{2}{\pi} \rho} \tr (\P^{\tran} \P^{*} \Psib^{\frac{1}{2}}) \bigg) \\
\label{eq:MSEp_blm_sls2} & = 1 \! + \! \frac{1}{K} \sum_{k=1}^{K} \bigg( \! (\rho K \! + \! 1) \Psi_{k} (\tau \! + \! \delta_{k}) \! - \! 2 \sqrt{\frac{2}{\pi} \Psi_{k} \rho} \tau \! \bigg).
\end{align}
Since each term in the summation of \eqref{eq:MSEp_blm_sls2} is a convex function of $\Psi_{k}$, the expression of the UE-specific scaling factor in \eqref{eq:Psi_k} can be obtained by setting ${\frac{\diff}{\diff \Psi_{k}} \eqref{eq:MSEp_blm_sls2} = 0}$. Finally, plugging \eqref{eq:Psi_k} into \eqref{eq:MSEp_blm_sls2} yields the MSE of the BLM~estimator~in~\eqref{eq:MSE_blm}. \hfill \IEEEQED

\begin{figure*}
\addtocounter{equation}{+17}
\begin{align}
\nonumber & \Exp \Big[ \sgn \Big( \sqrt{\rho} \big( \Re[h_{m}] \Re[p_{u}] + \Im[h_{m}] \Im[p_{u}] \big) + \Re[Z_{m,u}] \Big) \sgn \Big( \sqrt{\rho} \big( \Re[h_{m}] \Re[s_{\ell}] - \Im[h_{m}] \Im[s_{\ell}] \big) + \Re[z_{m}] \Big) \Big] \\
\label{eq:big_exp4} & = \Omega \bigg( \frac{\rho (\Re[p_{u}] \Re[s_{\ell}] - \Im[p_{u}] \Im[s_{\ell}])}{\sqrt{\rho (\Re[p_{u}]^{2} + \Im[p_{u}]^{2}) + 1} \sqrt{\rho (\Re[s_{\ell}]^{2} + \Im[s_{\ell}]^{2}) + 1}} \bigg)
\end{align}
\hrulefill
\addtocounter{equation}{+1}
\begin{align}
\nonumber |\El|^{2} & = (\rho + 1)^{2} \Psi^{\prime} M^{2} \sum_{u=1}^{\tau} \sum_{v=1}^{\tau} \bigg( \Re[p_{u}^{*} p_{v}] \bigg( \Omega \bigg( \frac{\rho \Re[p_{u} s_{\ell}]}{\sqrt{(\rho + 1)(\rho |s_{\ell}|^{2} + 1)}} \bigg) \Omega \bigg( \frac{\rho \Re[p_{v} s_{\ell}]}{\sqrt{(\rho + 1)(\rho |s_{\ell}|^{2} + 1)}} \bigg) \\
\nonumber & \phantom{=} \ + \Omega \bigg( \frac{\rho \Im[p_{u} s_{\ell}]}{\sqrt{(\rho + 1)(\rho |s_{\ell}|^{2} + 1)}} \bigg) \Omega \bigg( \frac{\rho \Im[p_{v} s_{\ell}]}{\sqrt{(\rho + 1)(\rho |s_{\ell}|^{2} + 1)}} \bigg) \bigg) + \Im[p_{u}^{*} p_{v}] \bigg( \Omega \bigg( \frac{\rho \Re[p_{u} s_{\ell}]}{\sqrt{(\rho + 1)(\rho |s_{\ell}|^{2} + 1)}} \bigg) \\
\label{eq:s_mean_abs2} & \phantom{=} \ \times \Omega \bigg( \frac{\rho \Im[p_{v} s_{\ell}]}{\sqrt{(\rho + 1)(\rho |s_{\ell}|^{2} + 1)}} \bigg) - \Omega \bigg( \frac{\rho \Im[p_{u} s_{\ell}]}{\sqrt{(\rho + 1)(\rho |s_{\ell}|^{2} + 1)}} \bigg) \Omega \bigg( \frac{\rho \Re[p_{v} s_{\ell}]}{\sqrt{(\rho + 1)(\rho |s_{\ell}|^{2} + 1)}} \bigg) \bigg) \bigg)
\end{align}
\hrulefill
\begin{align}
\label{eq:s_var_re1} \Re[\hat{s}_{\ell}]^{2} & = \frac{(\rho + 1)^{2}}{4} \Psi^{\prime} \bigg( \sum_{m=1}^{M} \sum_{u=1}^{\tau} \big( \Re[p_{u}] (a_{m,u} c_{m} + b_{m,u} d_{m}) + \Im[p_{u}] (a_{m,u} d_{m} - b_{m,u} c_{m}) \big) \bigg)^{2} \\
\nonumber & = \frac{(\rho + 1)^{2}}{4} \Psi^{\prime} \bigg( \sum_{m=1}^{M} \sum_{u=1}^{\tau} \big( \Re[p_{u}] (a_{m,u} c_{m} + b_{m,u} d_{m}) + \Im[p_{u}] (a_{m,u} d_{m} - b_{m,u} c_{m}) \big)^{2} \\
\nonumber & \phantom{=} \ + \sum_{m=1}^{M} \sum_{u \neq v} \big( \Re[p_{u}] (a_{m,u} c_{m} + b_{m,u} d_{m}) + \Im[p_{u}] (a_{m,u} d_{m} - b_{m,u} c_{m}) \big) \big( \Re[p_{v}] (a_{m,v} c_{m} + b_{m,v} d_{m}) \\
\nonumber & \phantom{=} \ + \Im[p_{v}] (a_{m,v} d_{m} - b_{m,v} c_{m}) \big) + \sum_{m \neq n} \sum_{u=1}^{\tau} \sum_{v=1}^{\tau} \big( \Re[p_{u}] (a_{m,u} c_{m} + b_{m,u} d_{m}) + \Im[p_{u}] (a_{m,u} d_{m} - b_{m,u} c_{m}) \big) \\
\label{eq:s_var_re2} & \phantom{=} \ \times \big( \Re[p_{v}] (a_{n,v} c_{n} + b_{n,v} d_{n}) + \Im[p_{v}] (a_{n,v} d_{n} - b_{n,v} c_{n}) \big) \bigg)
\end{align}
\hrulefill
\begin{align}
\nonumber |\hat{s}_{\ell}|^{2} & = \frac{(\rho + 1)^{2}}{4} \Psi^{\prime} \bigg( 4 M \tau + 2 \sum_{m=1}^{M} \sum_{u \neq v} \big( \Re[p_{u}^{*} p_{v}] (a_{m,u} a_{m,v} + b_{m,u} b_{m,v}) - \Im[p_{u}^{*} p_{v}] (a_{m,u} b_{m,v} - b_{m,u} a_{m,v}) \big) \\
\nonumber & \phantom{=} \ + \sum_{m \neq n} \sum_{u=1}^{\tau} \sum_{v=1}^{\tau} \Big( \Re[p_{u}^{*} p_{v}] \big( (a_{m,u} c_{m} + b_{m,u} d_{m}) (a_{n,v} c_{n} + b_{n,v} d_{n}) + (a_{m,u} d_{m} - b_{m,u} c_{m}) (a_{n,v} d_{n} - b_{n,v} c_{n}) \big) \\
\label{eq:s_abs2} & \phantom{=} \ + \Im[p_{u}^{*} p_{v}] \big( (a_{m,u} c_{m} + b_{m,u} d_{m}) (a_{n,v} d_{n} - b_{n,v} c_{n}) - (a_{m,u} d_{m} - b_{m,u} c_{m}) (a_{n,v} c_{n} + b_{n,v} d_{n}) \big) \Big) \bigg)
\end{align}
\hrulefill \vspace{-4mm}
\end{figure*}

\section{Proof of Theorem~\ref{thm:s_mean}} \label{app:s_mean}

We begin by introducing the following definitions:\footnote{Note that \eqref{eq:a1}--\eqref{eq:b2} are equivalent to \eqref{eq:A1}--\eqref{eq:B2} for $K=1$.}
\addtocounter{equation}{-23}
\begin{align}
\label{eq:a1} a_{m,u} & \triangleq \sgn \big( \Re[\sqrt{\rho} h_{m} p_{u}^{*} + Z_{m,u}] \big) \\
\nonumber & = \sgn \Big( \sqrt{\rho} \big( \Re[h_{m}] \Re[p_{u}] + \Im[h_{m}] \Im[p_{u}] \big) \\
\label{eq:a2} & \phantom{=} \ + \Re[Z_{m,u}] \Big), \\
\label{eq:b1} b_{m,u} & \triangleq \sgn \big( \Im[\sqrt{\rho} h_{m} p_{u}^{*} + Z_{m,u}] \big) \\
\nonumber & = \sgn \Big( \sqrt{\rho} \big( - \Re[h_{m}] \Im[p_{u}] + \Im[h_{m}] \Re[p_{u}] \big) \\
\label{eq:b2} & \phantom{=} \ + \Im[Z_{m,u}] \Big), \\
\label{eq:c1} c_{m} & \triangleq \sgn \big( \Re[\sqrt{\rho} h_{m} s_{\ell} + z_{m}] \big) \\
\label{eq:c2} & = \sgn \Big( \! \sqrt{\rho} \big( \Re[h_{m}] \Re[s_{\ell}] \! - \! \Im[h_{m}] \Im[s_{\ell}] \big) \! + \! \Re[z_{m}] \! \Big)
\end{align}

\noindent and
\begin{align}
\label{eq:d1} d_{m} & \triangleq \sgn \big( \Im[\sqrt{\rho} h_{m} s_{\ell} + z_{m}] \big) \\
\label{eq:d2} & = \sgn \Big( \! \sqrt{\rho} \big( \Re[h_{m}] \Im[s_{\ell}] \! + \! \Im[h_{m}] \Re[s_{\ell}] \big) \! + \! \Im[z_{m}] \! \Big).
\end{align}
From \eqref{eq:x_hat_mrc}, we can write the estimated symbol as
\begin{align}
\label{eq:s_exp1} \hspace{-2mm} \hat{s}_{\ell} & = \frac{\rho \! + \! 1}{2} \sqrt{\Psi^{\prime}} \sum_{m=1}^{M} \sum_{u=1}^{\tau} p_{u}^{*} (a_{m,u} \! + \! j \, b_{m,u})^{*} (c_{m} \! + \! j \, d_{m}) \\
\nonumber & = \frac{\rho + 1}{2} \sqrt{\Psi^{\prime}} \sum_{m=1}^{M} \sum_{u=1}^{\tau} p_{u}^{*} \big( a_{m,u} c_{m} + b_{m,u} d_{m} \\
\label{eq:s_exp2} & \phantom{=} \ + j \, (a_{m,u} d_{m} - b_{m,u} c_{m}) \big).
\end{align}
Hence, the expression in \eqref{eq:s_mean} results from
\begin{align}
\label{eq:a_c_exp1} \Exp [a_{m,u} c_{m}] & = \Exp [b_{m,u} d_{m}] \\
\label{eq:a_c_exp2} & = \Omega \bigg( \frac{\rho \Re[p_{u} s_{\ell}]}{\sqrt{(\rho + 1)(\rho |s_{\ell}|^{2} + 1)}} \bigg), \\
\label{eq:a_d_exp1} \Exp [a_{m,u} d_{m}] & = - \Exp [b_{m,u} c_{m}] \\
\label{eq:a_d_exp2} & = \Omega \bigg( \frac{\rho \Im[p_{u} s_{\ell}]}{\sqrt{(\rho + 1)(\rho |s_{\ell}|^{2} + 1)}} \bigg)
\end{align}
which are derived again by applying Proposition~\ref{pro:exp_sign}. For instance, $\Exp [a_{m,u} c_{m}]$ in \eqref{eq:a_c_exp1}--\eqref{eq:a_c_exp2} can be obtained by plugging
\begin{align}
\zetab & = \big[ \Re[h_{m}], \Im[h_{m}], \Re[Z_{m,u}], \Re[z_{m}] \big]^{\tran} \sim \setN \bigg( \0_{4}, \frac{1}{2} \I_{4} \bigg), \\
\a_{1} & = \big[ \sqrt{\rho} \Re[p_{u}], \sqrt{\rho} \Im[p_{u}], 1, 0 \big]^{\tran} \in \Real^{4 \times 1}, \\
\a_{2} & = \big[ \sqrt{\rho} \Re[s_{\ell}], - \sqrt{\rho} \Im[s_{\ell}], 0, 1 \big]^{\tran} \in \Real^{4 \times 1}
\end{align}
into \eqref{eq:exp_sign}, which gives \eqref{eq:big_exp4} at the top of the page. \hfill \IEEEQED

\begin{figure*}
\addtocounter{equation}{+6}
\begin{align}
\nonumber \Exp \big[ |\hat{s}_{\ell}|^{2} \big] & = (\rho + 1)^{2} \Psi^{\prime} M \bigg( \tau + \sum_{u \neq v} \bigg( \Re[p_{u}^{*} p_{v}] \Omega \bigg( \frac{\rho \Re[p_{u} p_{v}^{*}])}{\rho + 1} \bigg) - \Im[p_{u}^{*} p_{v}] \Omega \bigg( \frac{\rho \Im[p_{u} p_{v}^{*}])}{\rho + 1} \bigg) \bigg) \\
\nonumber & \phantom{=} \ + (M \! - \! 1) \sum_{u=1}^{\tau} \sum_{v=1}^{\tau} \Re[p_{u}^{*} p_{v}] \bigg( \! \Omega \bigg( \! \frac{\rho \Re[p_{u} s_{\ell}])}{\sqrt{(\rho \! + \! 1)(\rho |s_{\ell}|^{2} \! + \! 1)}} \! \bigg) \Omega \bigg( \! \frac{\rho \Re[p_{v} s_{\ell}])}{\sqrt{(\rho \! + \! 1)(\rho |s_{\ell}|^{2} \! + \! 1)}} \! \bigg) \! + \! \Omega \bigg( \! \frac{\rho \Im[p_{u} s_{\ell}])}{\sqrt{(\rho \! + \! 1)(\rho |s_{\ell}|^{2} \! + \! 1)}} \! \bigg) \\
\nonumber & \phantom{=} \ \times \Omega \bigg( \frac{\rho \Im[p_{v} s_{\ell}])}{\sqrt{(\rho + 1)(\rho |s_{\ell}|^{2} + 1)}} \bigg) \bigg) + (M - 1) \sum_{u=1}^{\tau} \sum_{v=1}^{\tau} \Im[p_{u}^{*} p_{v}] \bigg( \Omega \bigg( \frac{\rho \Re[p_{u} s_{\ell}])}{\sqrt{(\rho + 1)(\rho |s_{\ell}|^{2} + 1)}} \bigg) \\
\label{eq:s_abs2_exp2} & \phantom{=} \ \times \Omega \bigg( \frac{\rho \Im[p_{v} s_{\ell}])}{\sqrt{(\rho + 1)(\rho |s_{\ell}|^{2} + 1)}} \bigg) - \Omega \bigg( \frac{\rho \Im[p_{u} s_{\ell}])}{\sqrt{(\rho + 1)(\rho |s_{\ell}|^{2} + 1)}} \bigg) \Omega \bigg( \frac{\rho \Re[p_{v} s_{\ell}])}{\sqrt{(\rho + 1)(\rho |s_{\ell}|^{2} + 1)}} \bigg) \bigg) \bigg)
\end{align}
\hrulefill \vspace{-4mm}
\end{figure*}

\section{Proof of Theorem~\ref{thm:s_var}} \label{app:s_var}

The variance of the estimated symbol $\hat{s}_{\ell}$ can be written as
\addtocounter{equation}{-6}
\begin{align} \label{eq:s_var1}
\Vl & = \Exp \big[ |\hat{s}_{\ell}|^{2} \big] - |\El|^{2}
\end{align}
with $\Exp \big[ |\hat{s}_{\ell}|^{2} \big] = \Exp \big[ \Re[\hat{s}_{\ell}]^{2} \big] + \Exp \big[ \Im[\hat{s}_{\ell}]^{2} \big]$ and where $|\El|^{2}$ is given by \eqref{eq:s_mean_abs2} at the top of the page (cf. \eqref{eq:s_mean}). Furthermore, recalling the definitions in \eqref{eq:a1}--\eqref{eq:d2} and building on \eqref{eq:s_exp1}--\eqref{eq:s_exp2}, we can write $\Re[\hat{s}_{\ell}]^{2}$ as in \eqref{eq:s_var_re1}--\eqref{eq:s_var_re2} at the top of the page, and $\Im[\hat{s}_{\ell}]^{2}$ can be obtained following similar steps. Then, summing up \eqref{eq:s_var_re2} and $\Im[\hat{s}_{\ell}]^{2}$ yields $|\hat{s}_{\ell}|^{2}$ in \eqref{eq:s_abs2} at the top of the page. Now, we have that $\xi_{m,u} \in \{a_{m,u}, b_{m,u}, c_{m}, d_{m}\}$ and $\xi_{n,v} \in \{a_{n,v}, b_{n,v}, c_{n}, d_{n}\}$ are independent random variables if $m \neq n$ regardless of the indices $u,v$. This implies that $\Exp [a_{m,u} c_{m} b_{n,v} d_{n}] = \Exp [a_{m,u} c_{m}] \Exp [b_{n,v} d_{n}]$ and the same holds for the other products in the second summation of \eqref{eq:s_abs2}. Hence, the expected value of \eqref{eq:s_abs2} is given by \eqref{eq:s_abs2_exp2} at the top of the page, which follows from \eqref{eq:a_c_exp1}--\eqref{eq:a_d_exp2} and from\footnote{Note that \eqref{eq:a_a_exp1}--\eqref{eq:a_b_exp} are equivalent to \eqref{eq:A_A_exp1}--\eqref{eq:A_B_exp} for $K=1$.}
\addtocounter{equation}{+5}
\begin{align}
\label{eq:a_a_exp1} \Exp [a_{m,u} a_{m,v}] & = \Exp [b_{m,u} b_{m,v}] \\
\label{eq:a_a_exp2} & = \Omega \bigg( \frac{\rho \Re[p_{u} p_{v}^{*}]}{\rho + 1} \bigg), \\
\label{eq:a_b_exp} \Exp [a_{m,u} b_{m,v}] & = \Omega \bigg( \frac{\rho \Im[p_{u} p_{v}^{*}]}{\rho + 1} \bigg).
\end{align}
Finally, the expression in \eqref{eq:s_var} is obtained by plugging \eqref{eq:s_mean_abs2} and \eqref{eq:s_abs2_exp2} into \eqref{eq:s_var1}. \hfill \IEEEQED

\bibliographystyle{IEEEtran}
\bibliography{IEEEabbr,refs}

\end{document}

%% file: Definitions.tex
\renewcommand{\a}{\mathbf{a}}

\newcommand{\h}{\mathbf{h}}
\newcommand{\p}{\mathbf{p}}

\renewcommand{\r}{\mathbf{r}}

\newcommand{\x}{\mathbf{x}}
\newcommand{\y}{\mathbf{y}}
\newcommand{\z}{\mathbf{z}}

\newcommand{\0}{\mathbf{0}}
\newcommand{\1}{\mathbf{1}}

\newcommand{\A}{\mathbf{A}}

\renewcommand{\H}{\mathbf{H}}
\newcommand{\I}{\mathbf{I}}

\renewcommand{\P}{\mathbf{P}}

\newcommand{\R}{\mathbf{R}}

\newcommand{\V}{\mathbf{V}}

\newcommand{\Y}{\mathbf{Y}}
\newcommand{\Z}{\mathbf{Z}}

\newcommand{\zetab}{\boldsymbol{\zeta}}

\newcommand{\Sigmab}{\mathbf{\Sigma}}

\newcommand{\Phib}{\mathbf{\Phi}}
\newcommand{\Psib}{\mathbf{\Psi}}

\newcommand{\setC}{\mathcal{C}}

\newcommand{\setN}{\mathcal{N}}

\newcommand{\setQ}{\mathcal{Q}}

\newcommand{\setS}{\mathcal{S}}

\newcommand{\Real}{\mbox{$\mathbb{R}$}}
\newcommand{\Compl}{\mbox{$\mathbb{C}$}}
\newcommand{\rmF}{\mathrm{F}}

\newcommand{\Diag}{\mathrm{Diag}}
\newcommand{\diff}{\mathrm{d}}
\newcommand{\Exp}{\mathbb{E}}
\newcommand{\herm}{\mathrm{H}}
\renewcommand{\Im}{\mathrm{Im}}
\renewcommand{\Pr}{\mathbb{P}}

\renewcommand{\Re}{\mathrm{Re}}
\newcommand{\sgn}{\mathrm{sgn}}
\newcommand{\tr}{\mathrm{tr}}
\newcommand{\tran}{\mathrm{T}}
\newcommand{\Var}{\mathbb{V}}